\documentclass[aps,pra,twocolumn,reprint,noeprint,superscriptaddress,amsmath,amssymb]{revtex4-1}
\usepackage{graphicx}
\usepackage{bm}
\usepackage{color}
\usepackage{subfigure}
\usepackage{bbm}
\usepackage{amsmath}
\usepackage[colorlinks, linkcolor=blue, urlcolor=blue, citecolor=blue]{hyperref}

\begin{document}
\title{Quantum enhanced sensing by echoing spin-nematic squeezing in atomic Bose-Einstein condensate}
\author{Tian-Wei Mao}
\thanks{These authors contributed equally to this work.}
\affiliation{State Key Laboratory of Low Dimensional Quantum Physics, Department of Physics, Tsinghua University, Beijing 100084, China}
\author{Qi Liu}
\thanks{These authors contributed equally to this work.}
\affiliation{Laboratoire Kastler Brossel, Coll\`ege de France, CNRS, ENS-PSL University,\\
 Sorbonne Universit\'e, 11 Place Marcelin Berthelot, 75005 Paris, France}
\author{Xin-Wei Li}
\affiliation{Graduate School of China Academy of Engineering Physics, Beijing 100193, China}
\author{Jia-Hao Cao}
\affiliation{State Key Laboratory of Low Dimensional Quantum Physics, Department of Physics, Tsinghua University, Beijing 100084, China}
 \author{Feng Chen}
\affiliation{State Key Laboratory of Low Dimensional Quantum Physics, Department of Physics, Tsinghua University, Beijing 100084, China}
\author{Wen-Xin Xu}
\affiliation{State Key Laboratory of Low Dimensional Quantum Physics, Department of Physics, Tsinghua University, Beijing 100084, China}
\author{Meng Khoon Tey}
 \affiliation{State Key Laboratory of Low Dimensional Quantum Physics, Department of Physics, Tsinghua University, Beijing 100084, China}
 \affiliation{Frontier Science Center for Quantum Information, Beijing 100193, China}
 \affiliation{ Collaborative Innovation Center of Quantum Matter, Beijing 100084, China}
 \affiliation{Hefei National Laboratory, Hefei, Anhui 230088, China}
 \author{Yi-Xiao Huang}
\email[]{yxhuang@zust.edu.cn}
 \affiliation{School of Science, Zhejiang University of Science and Technology,\\
 Hangzhou, Zhejiang 310023, China}
\author{Li You}
\email[]{lyou@mail.tsinghua.edu.cn}
 \affiliation{State Key Laboratory of Low Dimensional Quantum Physics, Department of Physics, Tsinghua University, Beijing 100084, China}
 \affiliation{Frontier Science Center for Quantum Information, Beijing 100193, China}
 \affiliation{ Collaborative Innovation Center of Quantum Matter, Beijing 100084, China}
 \affiliation{Hefei National Laboratory, Hefei, Anhui 230088, China}
 \affiliation{Beijing Academy of Quantum Information Sciences, Beijing 100193, China}
\date{\today}

%*These authors contributed equally to this work.\\
%$\dagger$lyou@mail.tsinghua.edu.cn\par
%\textbf{}\par

\begin{abstract}
	Quantum entanglement can provide enhanced precision beyond standard quantum limit (SQL), the highest precision achievable with classical means\cite{Giovannetti:2004aa}. It remains challenging, however, to observe large enhancement limited by the experimental abilities to prepare, maintain, manipulate and detect entanglement\cite{Pezze2018}. Here, we present nonlinear interferometry protocols based on echoing spin-nematic squeezing to achieve record high enhancement factors in atomic Bose-Einstein condensate. The echo is realized by a state-flip of the spin-nematic squeezed vacuum, which serves as the probe state and is refocused back to the vicinity of the unsqueezed initial state while carrying out near noiseless amplification of a signal encoded. A sensitivity of $21.6\pm0.5$ decibels (dB) for a small-angle Rabi rotation beyond the two-mode SQL of 26400 atoms as well as $16.6\pm1.3$ dB for phase sensing in a Ramsey interferometer are observed. The absolute phase sensitivity for the latter extrapolates to $103~\rm{pT/\sqrt{Hz}}$ at a probe volume of $18~\mu\rm{m}^3$ for near-resonant microwave field sensing. Our work highlights the excellent many-body coherence of spin-nematic squeezing and suggests its possible quantum metrological applications in atomic magnetometer\cite{Muessel:2014aa, Vengalattore:2007aa, Yang:2020aa}, atomic optical clock\cite{Kruse, Pedrozo-Penafiel:2020aa}, and fundamental testing of Lorentz symmetry violation\cite{Dzuba:2016aa, Li:2019aa, zhuang}, etc.
\end{abstract}
\maketitle

%The widely discussed examples involve using squeezed light for (passive and linear) interferometers,
%like Michelson-Morley such as Mach-Zehnder, Fabry-Perot
%with quantum entangled states such as squeezed states of photons, spin squeezed states of atoms, or Dicke states.
%The with . and extensively demonstrated
% help achieve be used to   \\
%
%
%SU(1,1) interferometer(SUI), which is characteristic of an entangling - phase encoding - disentangling process,  has long been proposed as an active device able to boost the phase sensitivity over classical limit. However since the particles inside SUI is typically much fewer than those in the pump modes, the resulting poorly \textit{absolute} phase sensitivity limits the practical use of conventional SUI.  Here we enhance its performance by heavily out-coupling atoms from pump mode to side-modes in a cloud of spinor Bose-Einstein condensate. The second step of spin-mixing dynamics servers as an interaction-based readout which translates the encoded phase to the number of atoms in side-modes. In experiments, we directly Our work enriches the toolbox of quantum enhanced metrology, and opens the door of exploring active interferometers in other nonlinear systems.
%
%\section{Introduction}
Using superposition states for quantum science and technology inevitably encounters projection noise\cite{Wineland:1992vb}, the indeterminacy of measurement outcome that erects a lower bounded precision based on statistical inference of independent measurements. This bound, the standard quantum limit (SQL) or classical precision limit, is given by $1/({M}-1)\sqrt{N}$ for $N$ uncorrelated particles in $M$-mode (path) interferometry\cite{Giovannetti:2004aa, Zou6381}. Correlated measurement outcomes from entangled particles can give rise to reduced uncertainty, leading to enhanced signal-to-noise ratio (SNR), or quantum enhanced precision\cite{Pezze2018}. Advancing such quantum enhancement is a frontier topic of research, which becomes especially valuable when the numbers of interfering particles are finite, e.g., photons in gravitational wave detection\cite{LIGO_squeezed} or bio-materials imaging\cite{Taylor:2016uh}, atoms or trapped ions in clocks\cite{Ludlow:2015ul,Pedrozo-Penafiel:2020aa,Burt:2021vj} or electromagnetic field sensing\cite{kevin,Bao2020}.

Studies on quantum enhanced precision with neutral atoms have progressed significantly after cold atoms or even quantum degenerate ensembles are employed. Quantum non-demolition measurement provides noise reduction in cold thermal atomic clouds\cite{Hosten2016, Bao2020}, and coherent spin dynamics in Bose-Einstein condensate (BEC) generates multi-atom entangled state with squeezed quantum noise\cite{Gross2010, Riedel2010, Hamley2012, Strobel:2014ux, Twin_Matter, Luo620, Zou6381}.
%it is still challenging to fully exploit entanglement resources and saturate the ultimate measurment precision given by quantum Cramer-Rao bound\cite{Helstrom:1967aa, Braunstein:1994aa}. 
Their observed precision enhancement factors remain limited by decoherence and dissipation, or detection noise from imperfect counting of particles. 
Recently, nonlinear interferometries with interaction-based readout protocols are proposed for achieving Heisenberg limited precision without requiring single particle resolution\cite{Loschmidt_echo,Gabbrielli2015, Davis, Frowis2016, Szigeti, Nolan, TACT_magnification, TNT_magnification, IBR_spin, Huang:2018ur}. They have led to the observations of remarkable enhancements in small systems such as  phonon states of a single trapped ion\cite{Burd} or about 400 cold thermal atoms in an optical cavity\cite{Colombo:2022tz}. However, for larger systems with thousands or more atoms, enhancement factors reported in nonlinear interferometries stay modest ($<$ 8 dB) due to the difficulties of realizing  perfect time-reversed dynamics from sign-flipping many-body interaction or from maintaining coherent long-term nonlinear dynamics\cite{Hosten,FirstTRexp, Liu2021}.

\begin{figure*}[t] 
	\centering
	\includegraphics[width= 0.95\linewidth]{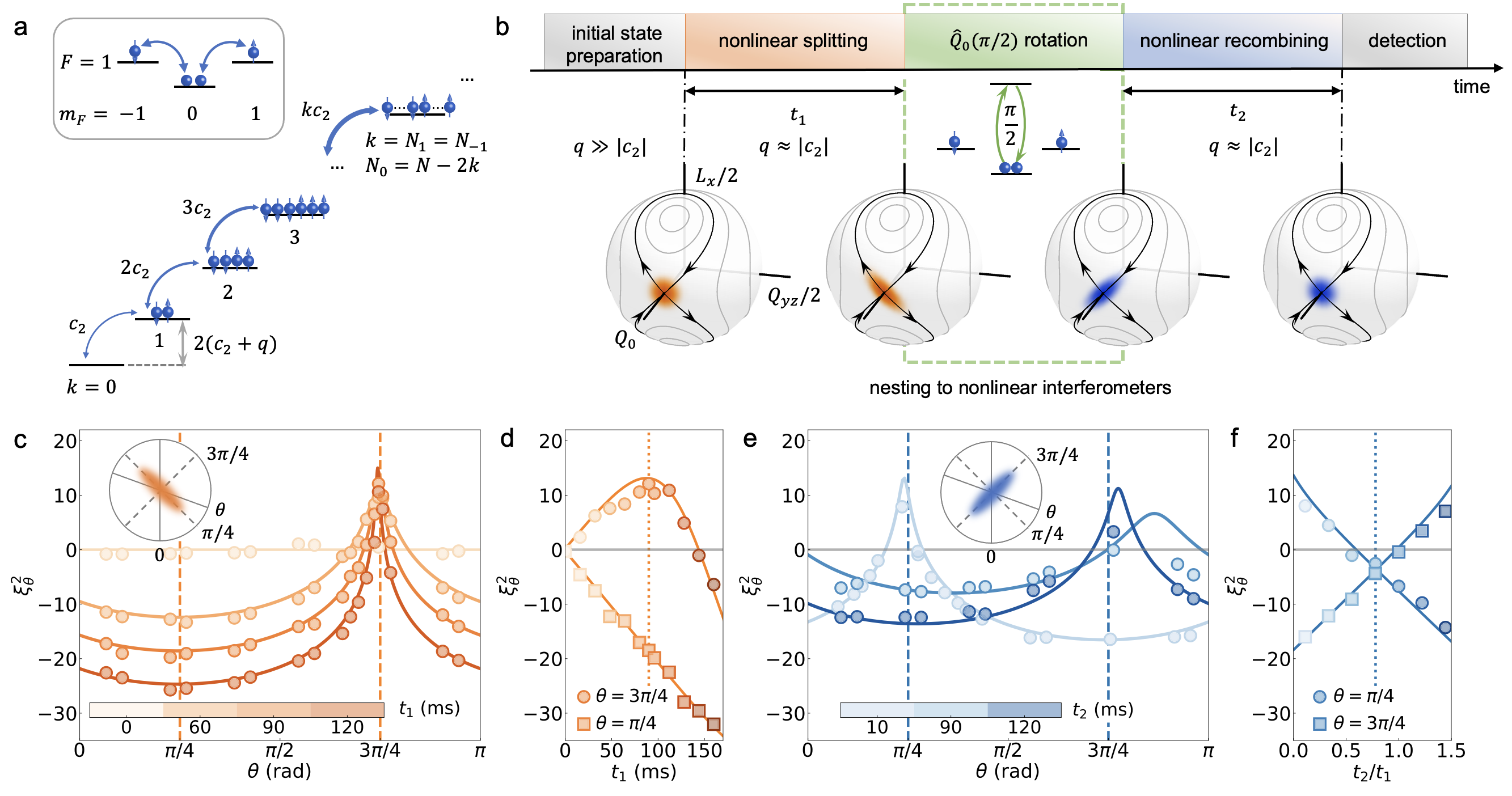}
	\caption{{\bf Spin-nematic squeezing nonlinear atom interferometry.}  {\bf a}, The enclosed upper part illustrates the level diagram for spin-1 atom, neglecting linear Zeeman shift. The Fock state basis $\left|k\right\rangle$ of $k$ atom pairs in $\left|\pm1\right\rangle$ shown in the lower part forms an approximate decoherence-free subspace, where all states within possess the same first order Zeeman shift and mutual coherence are insensitive to magnetic field noise. The increasing coupling strength (denoted by thicker double arrowed lines) and the constant energy gap between adjacent states $\left|k\right\rangle$ and $\left|k-1\right\rangle$ are given by $kc_2$ and $2(c_2+q)$ respectively. {\bf b}, Two sections of spin-nematic squeezing (lasting  respectively for $t_1$ and $t_2$) at $q\approx|c_2|$ form the basic elements for the discussed nonlinear interferometry, preceded with  state initialization and followed by state detection. The state-flip by $\pi/2$ rotation around $\hat Q_0$ comes from the relative phase between two microwave pulses that respectively drive atoms to and back from an ancillary state ($\left|F=2,m_F=0\right\rangle$), making the subsequent dynamics effectively time-reversed as in an echo, refocusing the state to the initial one. This rotation or state-flip will be nested with encoding operations in Figs.~2-3 respectively for sensing RF rotation and quadrature phase. Numerically simulated state distributions are shown on the spin-nematic spheres for ${N = 100}$ atoms.  
	{\bf c}, Measured spin-nematic squeezing in the $\hat L_x$-$\hat Q_{yz}$ plane. {\bf d}, The squeezing parameter along quadrature angle $\theta=3\pi/4$ for optimal squeezing and $\pi/4$ (optimal anti-squeezing) versus evolution time. {\bf e}, The squeezing parameter of the effectively time-reversed dynamics after encoding operation  at $t_1=90$ ms (denoted by the dotted line in {\bf d}), and {\bf f} the dependencies of squeezing parameter at $\theta=\pi/4$ and $3\pi/4$ on $t_2$ show optimal unsqueezing  at $t_2/t_1=0.78$ (blue dotted line). In all panels {\bf c-f}, markers denote experimental results averaged over 100 runs and error bars are smaller than the marker sizes. Solid lines denote semi-classical calculations based on truncated Wigner approximation (more details can be found in SI).}
	\label{fig:schematic}
\end{figure*}
Here, we report record high quantum enhanced metrological gains with nonlinear interferometries based on echoing spin-nematic squeezing in a BEC of 26400 $^{87}$Rb atoms. Coherent spin-mixing dynamics, which is responsible for spin-nematic squeezing, is used to construct the two essential elements of nonlinear interferometry: {\it nonlinear splitting} which generates spin-nematic squeezed vacuum\cite{two_spin_squeezing, Hamley2012} for probing, and {\it nonlinear recombining} which disentangles (unsqueezes) the probe state back into the vicinity of the (uncorrelated) initial state for nearly noiseless signal amplification. The latter operation is realized by a probe state-flip from switching the squeezing and anti-squeezing axes of the nematic squeezed vacuum, leading to effectively time-reversed dynamics, or echo of spin-nematic squeezing, reminiscent of the spin-squeezing echo\cite{Davis, Hosten} or the SU(1,1) echo\cite{FirstTRexp} with the state refocused back to the initial one. The well-known spin-echo refocuses state from effectively time-reversed linear dynamics under inhomogeneous coupling field by a $\pi$-pulse or state-flip, while in the present case, flipping of the probe state refocuses the nonlinear dynamics of spin-nematic squeezing.
We then complete the nonlinear interferometers by nesting the probe state encoding operations with specific sensing applications: a small-angle radio-frequency (RF) Rabi rotation, and a quadrature phase interrogation in microwave (MW) sensing Ramsey interferometry\cite{Szigeti}. The respectively achieved metrological gains reach $21.6\pm0.5$ dB and $16.6\pm1.3$ dB beyond the two-mode SQL, beating both previous records in any system to date\cite{Hosten2016,Colombo:2022tz}.
\begin{figure*}[htb]
	\centering
	\includegraphics[width= 1 \linewidth]{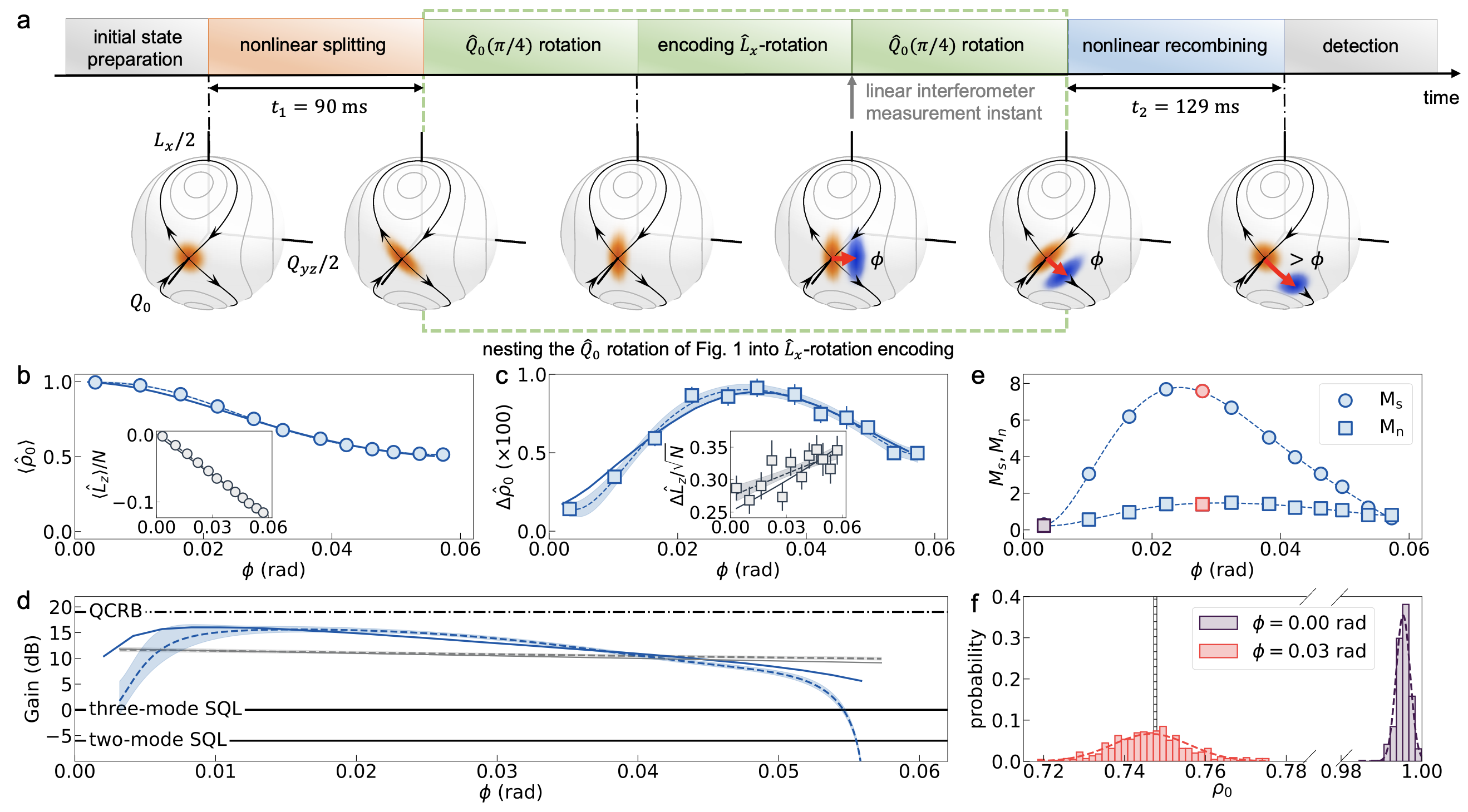}
	\caption{{\bf Nonlinear Rabi interferometry for sensing $\hat L_x$-rotation. }{\bf a}, Schematic of the Rabi interferometry protocol with {\it splitting} and {\it recombining} the same as in Fig.~\ref{fig:schematic}. The sequence inside the green dashed-line rectangular box details the nested operations of the state-flip in Fig.~\ref{fig:schematic}{\bf a}, making it a Rabi interferometer. The blue shaded state distribution on the spin-nematic sphere corresponds to the encoded state after Rabi rotation around $\hat L_x$ by an angle $\phi$ denoted by the short red line segment, which is magnified and shown elongated in the next sphere after time-reversed nonlinear recombining.  {\bf b-c}, The average and standard deviation of $\hat\rho_0$ at variable $\phi$. The insets present $\langle\hat L_z\rangle$ and $\Delta\hat L_z$ from linear interferometry with each point representing average of 100 repeated experiments. {\bf d}, Metrological gains of the nonlinear (blue dashed line) and linear interferometers (grey dashed line) obtained from error propagation. The dash-dotted line corresponds to quantum Cram\'er-Rao bound (QCRB), and the black solid lines stand for two or three-mode SQL. {\bf e}, Circles and squares denote the magnification factors respectively for slope ($M_s$) and noise ($M_n$) over linear interferometry. With increasing $\phi$, $M_s$ quickly ascends while $M_n$ remains essentially unchanged, leading to enhanced SNR over three-mode SQL due to signal (slope) amplification. {\bf f}, The Gaussian probability distributions for the final states of the nonlinear interferometer sampled over 500 times at $\phi= 0$ {rad} and $0.03$ {rad} respectively. The much broader distribution  of $\hat\rho_0$ than the atom number counting resolution (illustrated by the narrow black dashed bar) demonstrates high robustness to detection noise. In all panels, solid lines are simulation results and dashed lines are fitted curves to experimental data.}
	\label{fig:data1}
\end{figure*}

The system we consider is a spin-1 atomic BEC in the ${F}=1$ hyperfine ground state with Zeeman component $\left|m_F\right\rangle$ ($m_F=0,\pm1$). Assuming the spatial density profiles for all three components are the same\cite{Law,Yi}, the dynamics for the collective spin $\hat{\mathbf{L}}=(\hat L_x$, $\hat L_y$, $\hat L_z)$, where $\hat L_k=\sum_{j=1}^N\hat l_k^{(j)}$ with $\hat l_k^{(j)}$ the spin component of the $j$th atom, is described by the Hamiltonian
\begin{eqnarray}
%\d{\hat H{\rm{ = }}\frac{{{c_2}}}{{2{N_t}}}(\hat L_x^2 + \hat L_y^2 + \hat L_z^2 - 2{\hat N_t}) - \frac{q}{2}{\hat Q_0},}
\hat H{\rm{ = }}\frac{{{c_2}}}{{2{N}}}\hat{\mathbf{L}}^2 - q{\hat N_0},
%\d{\hat H{\rm{ = }}\frac{{{c_2}}}{{2{N_t}}}(\hat L_x^2 + \hat L_y^2 + \hat L_z^2 - 2{\hat N_t}) - \frac{q}{2}{\hat Q_0}}.
\label{Ham}
\end{eqnarray}
where $c_2$ and $N$ in the first term denote strength of spin-exchange interaction and total atom number respectively. The second term is proportional to atomic number operator $\hat N_0$ of $|0\rangle$ component, its coefficient $q$ measures quadratic Zeeman shift (QZS), which is tunable by a bias magnetic field or an off-resonant MW dressing field\cite{Gerbier,Jiang}. Figure~\ref{fig:schematic}{\bf a} illustrates the BEC system subspace with $k$ atom pairs in $\left|\pm1\right\rangle$ components, spanned by spin-mixing that transfers atoms from $|0\rangle$ to $\left|\pm1\right\rangle$ or vice versa when dynamic instability arises  for $q\in(0, 2|c_2|)$. It forms an approximate decoherence-free subspace, in which all states are of the same magnetization and quantum coherence within  are immune to first-order magnetic field noise. During short-term spin-mixing when the total number of paired atoms in $\left|\pm1\right\rangle$ is 
small, i.e., $\langle k\rangle\ll\langle\hat N_0\rangle$, the distribution of $k$ resembles that of the two-mode squeezed vacuum, the state is then appropriately called spin-nematic squeezed vacuum\cite{two_spin_squeezing,Hamley2012}. The nonlinear squeezing dynamics can be illustrated by the changing quasi-probability distribution on spin-nematic sphere, a sub-representation of spin-1 analogous to the Bloch sphere for spin-1/2 (see Supplementary Information (SI)), whose three axes $\left\{\hat Q_{yz}/2,\hat L_x/2,\hat Q_0\equiv(\hat Q_{yy}-\hat Q_{zz})/4\right\}$ span a SU(2) subgroup with $\hat Q_{ij}=\hat L_i\hat L_j+\hat L_j\hat L_i-(4/3)\delta_{ij}$ the associated quadrupole or nematic tensor operators of the full SU(3) group.

The {\it splitting} and {\it recombining} operations of the basic nonlinear interferometry protocol are shown in Fig.~1{\bf b}, both resulting from time-forward spin-mixing at $q\approx|c_2|$ governed by Eq.~(\ref{Ham}). The solid lines on each spin-nematic sphere denote contours of fixed mean field energy, among which the separatrix (dark black line) intersects perpendicularly at angle $\pi/2$ when $q=|c_2|$. The initial classical state with isotropic  Gaussian  distribution is observed to expand along one branch of the separatrix while shrinks along the other, leading to a squeezed distribution implicating quantum entanglement\cite{Hamley2012} as shown in Fig. \ref{fig:schematic}{\bf b}. If the {squeezed} state is flipped by  instantaneously rotating $\rm{\pi}/2 $ around $\hat Q_0$ axis to effectively interchange the squeezing and anti-squeezing directions, subsequent spin-mixing carries out effectively time-reversed dynamics which disentangles the state back to the initial polar state (with all atoms in $\left|0\right\rangle$) in the end. Such an induced echo of spin-nematic squeezing extends the nonlinear dynamics echo of atomic SU(1,1) interferometry explored earlier\cite{FirstTRexp}. 

We first characterize the dynamic evolutions for {\it nonlinear splitting} and {\it recombining} operations in a BEC of $N=26400\pm240$ atoms, which is prepared initially in the polar state under a bias magnetic field of 0.537 G inside a crossed optical dipole trap with tight harmonic trapping frequencies $2\pi\times(190,93,121)$ Hz. Spin-nematic squeezing is initiated by quenching QZS to $q\approx|c_2|=2\pi\times3.8$ Hz through turning on a dressing MW $42.33$-MHz blue detuned from the $|F = 1,m_F = 0\rangle$ to $|2,0\rangle$ clock transition and lasts for $t_1$. This is followed by spin-nematic unsqueezing for $t_2$ after state-slip by  rotating $\pi/2$ around $\hat Q_0$ axis. As illustrated below the time sequence axis of Fig.~\ref{fig:schematic}{\bf b}, $\hat Q_0(\pi/2)$ is implemented by two consecutive  resonant $\pi$ pulses with a $\pi/2$ relative phase that transfer atoms from $|1, 0\rangle $ to $| 2, 0\rangle $ and back in rapid succession (see Methods). The evolving anisotropic probability distributions during {\it splitting} and {\it recombining} operations are characterized in the spin-nematic plane by quadrature fluctuation operator $\hat Q(\theta)=\hat Q_{yz} \sin\theta- \hat L_x \cos\theta$, which is first mapped to magnetization $\hat L_z=\hat N_1-\hat N_{-1}$ followed by Stern-Gelach absorption imaging after 10 ms time-of-flight. The mapping is carried out by two rotations: a rotation of $\theta$ around $\hat Q_0$ axis to align $\hat Q(\theta)$ along the $\hat L_x$ axis, followed by a $\pi/2$ RF pulse rotation around $\hat L_y$.

The build-up of spin-nematic squeezing during {\it nonlinear splitting} is shown in Figs.~\ref{fig:schematic}{\bf c} and \ref{fig:schematic}{\bf d}, quantified by the squeezing parameter $\xi^2_{\theta} = -20{\log _{10}}(\Delta {\hat L_z}/\sqrt N )$, which vanishes at all quadrature phase $\theta$ for the polar state ($t_1=0$). The distribution becomes increasingly anisotropic for $t_1>0$ as a result of spin-nematic squeezing, leading to enlarged variations of the squeezing parameter with peak and dip around $\theta=3\pi/4$ and $\theta=\pi/4$ respectively, as observed in earlier pioneering studies\cite{Hamley2012}. The measured results agree well with semi-classical calculations (solid lines) based on truncated Wigner method taking into account atom loss and fluctuation of QZS (see SI). 
%For phases in the vicinity of $\theta=3\pi/4$ (Fig.~\ref{fig:schematic}{\bf d}), the relatively large discrepancy can be ascribed to the detection noise in the imaging system and the small calibration error of $q$ which slightly shifts the optimal squeezed axis. 
The best squeezing we observe is $12.1\pm0.5$ dB at $t_1=90$ ms, which improves to $14.0\pm0.5$ dB after subtracting the independently calibrated detection noise of $\Delta\hat L_z^{\rm{DN}} \approx 24$. For longer evolution time when the system reaches over-squeezed regime, both the measured squeezing and anti-squeezing are found to decline, due to the occurrence of non-Gaussian entanglement whose properties can no longer be faithfully represented by the squeezing parameter\cite{Strobel:2014ux} alone.\par

The effectively time-reversed dynamics after {$\hat Q_0(\pi/2)$} state-flip at $t_1=90$ ms is shown in Figs.~\ref{fig:schematic}{\bf e} and \ref{fig:schematic}{\bf f}. The peak-to-peak variation of the squeezing parameter $\xi^2_{\theta}$ initially decreases with increasing $t_2$, consistent with echoed dynamics towards the initial state. The evolution of squeezing and anti-squeezing at $\theta=\pi/4$ and $\theta=3\pi/4$ (interchanged after {$\hat Q_0(\pi/2)$ }) are recorded in Fig.~\ref{fig:schematic}{\bf f}. Their difference reaches a minimum at $t_2=70$ ms when the observed effectively time-reversed evolution best disentangles the probe state.  This instant is slightly ahead of the expected value $t_2=t_1$, which can be ascribed to the small discrepancy between $q$ and $|c_2|$. With further increasing $t_2$, squeezing and its associated anti-squeezing build up again, accompanied by a second switch of their respective optimal angles, and continue on to repeat the first stage dynamics shown in Fig.~\ref{fig:schematic}{\bf c}.

Previous studies of spin-nematic squeezing reported 8-10 dB of quantum noise suppression, limited mainly by {dissipation and} detection noise\cite{Hamley2012}.
{Here, with an order of magnitude longer condensate lifetime and high spin coherence demonstrated in the above calibrations}, complete spin-nematic nonlinear atom interferometry is implemented for the first time, parallel to related theoretical developments\cite{Szigeti, Niezgoda_2019}. As shown inside the green dashed-line rectangular box in Fig.~\ref{fig:data1}{\bf a}, by separating the {$\hat Q_0(\pi/2)$} of Fig.~\ref{fig:schematic}{\bf b} into two equal halves and augmenting an encoding rotation by $\phi$ around $\hat L_x$ as sandwiched in between, we arrive at the first nonlinear interferometer for sensing small-angle Rabi rotation $\phi$. The {$\phi$-dependent }fractional population $\hat\rho_0= \hat N_0/N$ is chosen as the estimator because it commutes with $\hat L_z$ and is immune to first-order magnetic field noise. We adopt the numerically optimized values for $t_1=90$ ms and $t_2=129$ ms. The unbalanced choice of $t_2>t_1$ is taken for further improved robustness to detection noise\cite{Nolan} (see SI). For comparison, a linear interferometer using the same spin-nematic squeezed vacuum probe state is carried out as well before preceding to {\it nonlinear recombining}. Its metrological performance is quantitatively characterized by projective measurement along the squeezed direction{, namely along the $\hat{Q}_{yz}$ axis} at the instant marked by the grey arrow in Fig.~\ref{fig:data1}{\bf a}.

Figures~\ref{fig:data1}{\bf {b-c}} present the $\phi$-dependent mean value and standard deviation of $\hat\rho_0$ ($\hat L_z$) for the discussed nonlinear (linear) Rabi interferometry. The metrological gain $-20\log_{10}\left[\Delta\phi/(\Delta\phi)_{\rm{SQL}}\right]$ is extracted with $\Delta\phi$ obtained from error propagation based on fitting measured data, and $(\Delta\phi)_{\rm{SQL}}=1/2\sqrt{N}$ denotes the three-mode SQL (see Methods). Both interferometers realize quantum enhanced gains {within the small rotation angle regime} as shown in Fig.~\ref{fig:data1}{\bf d}. The nonlinear Rabi interferometer reaches an optimal gain of $15.6\pm0.5$ dB, which translates to $21.6\pm0.5$ dB with respect to the two-mode SQL of $1/\sqrt{N}$, outperforming the previous record of $18.5\pm0.3$ dB achieved in a cold thermal ensemble of about 0.7 million atoms\cite{Hosten2016}. It is interesting to note the observed  gain is higher than both the value of $11.7\pm0.3$ dB (or $17.7\pm0.3$ dB with respect to two-mode SQL) obtained in linear interferometry despite of the same spin-nematic squeezed vacuum used. 
%Surprisingly, this value is even higher than the directly measured squeezing parameter of $12.1\pm0.5$ dB {in Fig.~1\bf c} for the probe state. 

To further elucidate the enhanced metrological gain and the {principle} of spin-nematic nonlinear interferometry, we define magnification factor $M_s$ for signal (analogously $M_n$ for noise) as the ratio of the observed slope (fluctuation) to that of linear interferometry with atomic coherent state. {As shown in Fig.~2{\bf e}, }although $M_s$ stays close to zero in the limit of $\phi\rightarrow0$, it rapidly ascends with increasing small $\phi$ while $M_n$ remains essentially unaltered. As a result, the SNR $M_s/M_n$, hence the metrological gain, increases beyond three-mode SQL ($M_s/M_n=1$) as soon as $\phi>0$. However, at larger rotation angles, SNR falls off due to depleting $|0\rangle$ mode population. We also note in Fig.~\ref{fig:data1}{\bf f} for angles around the maximal SNR, the distribution of $\hat\rho_0$ becomes much broader than atom number counting  resolution (illustrated by the black dashed bar), making the  nonlinear Rabi interferometer robust to detection noise (see SI for detailed comparisons between the linear and nonlinear interferometers).

\begin{figure}[t]
	\centering
	\includegraphics[width=1\linewidth]{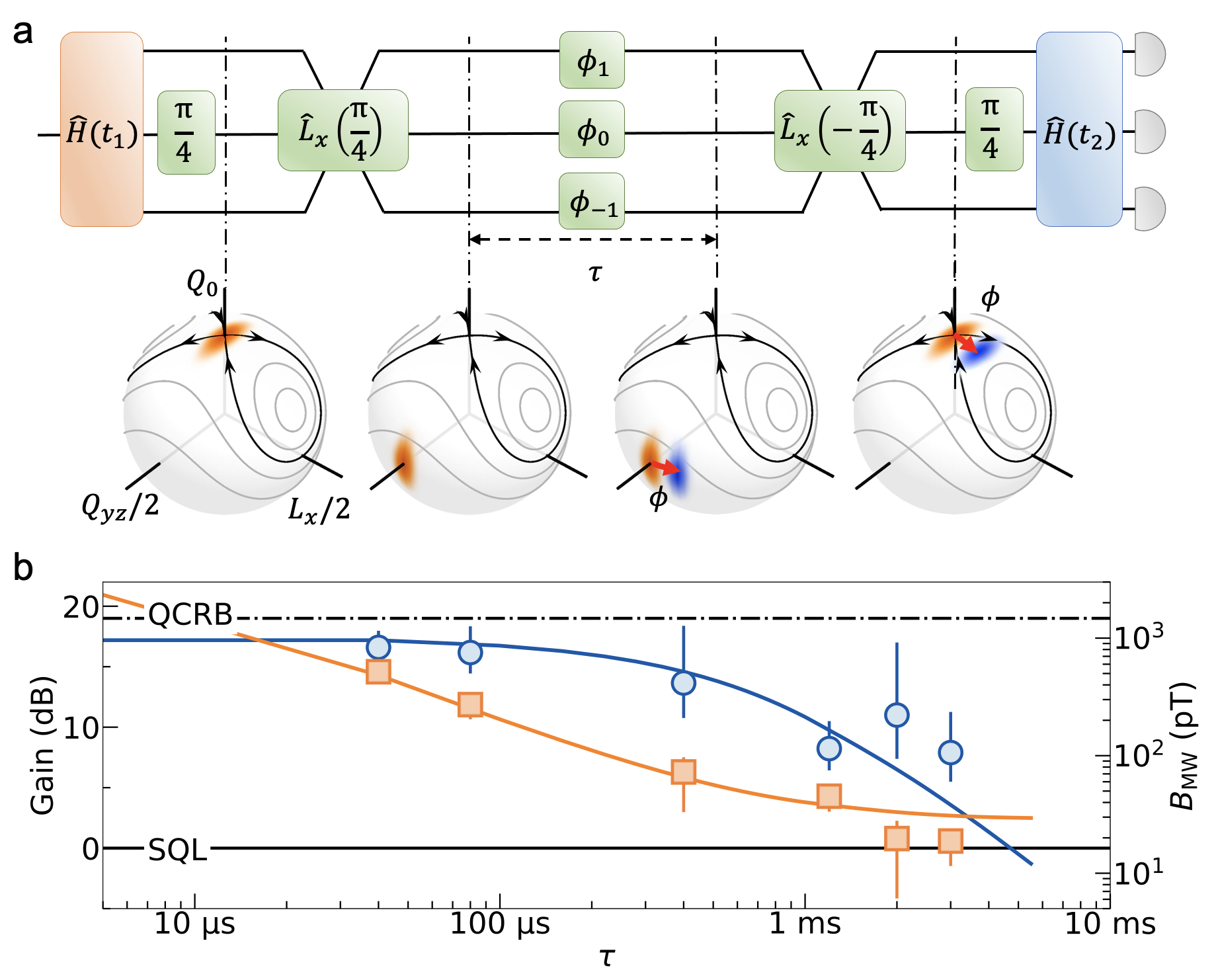}
	\caption{{\bf Nonlinear Ramsey interferometry for quadrature phase sensing. } {\bf a}, Schematic for the Ramsey interferometry protocol. Spin-nematic squeezed vacuum is rotated by $\hat Q_0(\pi/4)$ to align the anti-squeezing direction with $\hat Q_{yz}$ axis. A pair of RF $\hat L_x(\pm\pi/4)$ pulses rotate the state to and back from the spin-nematic sphere equator, allowing the quadrature phase $\phi=(\phi_1+\phi_{-1})/2-\phi_0$ to be interrogated from MW dressing. The second {$\hat Q_0(\pi/4)$ rotation} and {\it nonlinear recombining} map the phase to fractional population $\hat\rho_0$ for detection in the end. The complete operations are identical to Ramsey interferometry when viewed over the Bloch sphere, except now with entangling dynamics during the {\it nonlinear splitter} and disentangling dynamics in the {\it nonlinear recombiner}. {\bf b}, The dependence of the observed metrological gain (blue circles, referenced to the left vertical axis) and single-shot sensitivity of MW magnetic field (orange squares, referenced to the right vertical axis) on interrogation time $\tau$. Solid line denotes the numerical simulation results including  MW power fluctuation. Each data point is the fitted result of 1400 measurements, and error bars indicate statistical uncertainties,and $t_1=90$ ms and $t_2=129$ ms, the same as in Fig.~2.}
	\label{fig:data3}
\end{figure}

The above $\hat L_x$-rotation nonlinear Rabi interferometer can be expanded further to construct a `pumped-up' SU(1,1) interferometer\cite{Szigeti} for sensing quadrature phase $\phi=(\phi_1+\phi_{-1})/2-\phi_0$, by directly interrogating from a rotation around $\hat{Q}_0$. The corresponding nonlinear interferometry is constructed by replacing the {$\hat Q_0(\pi/2)$ state-flip rotation} in Fig.~1{\bf b} with a sequence of rotations as illustrated in Fig.~\ref{fig:data3}{\bf a}, to ensure the optimal squeezed direction is aligned orthogonal to the generator $\hat Q_0$ axis during interrogation. 
We apply this nonlinear Ramsey interferometry  to high-spatial-resolution sensing of microwave field (Fig.~\ref{fig:data3}{\bf b}), which is power stabilized and off-resonant from the $6.8$ GHz clock transition to encode a quadrature phase of $\phi=q\tau$ over interrogation time $\tau$. A RF spin-echo $\pi$-pulse is applied in the middle to further suppress magnetic field noise. We obtain a maximal phase sensitivity gain of $16.6\pm1.3$ dB, which outperforms the highest enhancement recently achieved in cold thermal atoms by $4.7$ dB\cite{Colombo:2022tz}. The record setting observations we report reflect the high quality of our pulse sequence control and the high coherence of many-body quantum entanglement during spin-nematic squeezing and its echo dynamics (for unsqueezing). The high level of quantum enhancement factor is maintained upto $\tau\sim3$ ms, limited by the fluctuation of $q$ due to MW instability in the present system. Our observation implies a $18.8$ pT single-shot sensitivity for near-resonant microwave magnetic field analogously defined (see SI). While at the same level as reported in a two-component atomic BEC\cite{Ockeloen:2013aa}, it is reached here within nearly one order of magnitude shorter interrogation time due to the significantly enhanced SNR. Taking the experimental cycle of $30$ s into consideration, our result implies a single shot MW magnetic field sensitivity of $103~\rm{pT/\sqrt{\rm{Hz}}}$, within a small probe of $18~\mu \rm{m}^3$ in volume.

In conclusion, we propose and implement two nonlinear interferometers by echoing spin-nematic squeezing in an atomic condensate. Record performances are observed for sensing a small-angle Rabi rotation 21.6 (15.6) dB beyond the two (three)-mode SQL of 26400 atoms, and 16.6 dB for sensing the quadrature phase from a power stabilized MW. Our nonlinear interferometry protocols employ state-flip induced effective time-reversal to overcome detection noise, mitigating the difficult  sign-flip of an interacting many-body Hamiltonian. They improve upon the pioneering atomic SU(1,1) interferometry\cite{FirstTRexp, Linnemann2017} {with an efficient exploitation of entangled probe state}, and they outperform the recently presented cyclic nonlinear interferometer\cite{Liu2021} by the reduced evolution time required{ for recurrence back to the initial state (see SI for detailed comparisons). Our studies thus open the door for developing quantum metrological applications with atomic spin-nematic squeezed vacuum, which exhibit significant practical advantages afforded by the inherent two-fold improved SQL due to “three-path” interference, robustness to noisy magnetic field from the decoherence free subspace and insensitivity to atom loss in the classical pump mode, etc.
% in spatial-resolved quantum magnetometry\cite{Vengalattore:2007aa,Yang:2020aa}, atomic optical clock for time-keeping\cite{Pedrozo-Penafiel:2020aa}, momentum entangled interferometer for sensing acceleration field or gravitational wave\cite{Anders:2021aa,Greve:2022aa}, and distributed quantum sensing network\cite{Malia}, etc.
It is also interesting to consider improving absolute field sensing sensitivity with the recently developed continuous BEC\cite{Chen:2022uz}. This could lead to drastically shortened experimental cycle, potentially overcoming a major bottle-neck for practical quantum metrology with ultracold atoms.

\bibliography{v7}

~\\
\textbf{Data availability} All data that support the plots within this paper and other findings of this study are available from the corresponding author upon reasonable request.\par
~\\
\textbf{Code availability}
All relevant codes or algorithms are available from the corresponding author upon reasonable request.\par
~\\
\textbf{Acknowledgements}
We thank Drs L. N. Wu, Y. Q. Zou, X. Y. Luo, J. L. Yu, M. Xue and S. F. Guo for helpful discussions. This work is supported by the National Natural Science
Foundation of China (NSFC) (Grants  No. 11654001 and No. U1930201), by Natural Science Foundation of Zhejiang Province (Grant No. LY22A050002), by the Key-Area Research and Development Program of GuangDong Province
(Grant No. 2019B030330001), and by the National Key R\&D Program of China (Grants No. 2018YFA0306504 and No. 2018YFA0306503).\par
~\\
\textbf{Author contributions}
Y.-X.H., Q.L., and L.Y. conceived this study. T.-W.M., Q.L., J.-H.C, and W.-X.X., performed the experiment and analyzed the data. T.-W.M., Q.L., X.-W.L., and F.C. conducted numerical simulations. T.-W.M., Q.L., Y.-X.H., M.-K.T. and L.Y. wrote the paper.\par
~\\
\textbf{Competing interests} The authors declare no competing interests.\par

\clearpage

\section*{Methods}
\setcounter{figure}{0}
\renewcommand{\figurename}{Extended Data Fig.}

\subsubsection*{Initial state preparation}
We initially prepare a BEC of about 27000 ${}^{87}$Rb atoms in $|F=1,m_F=0\rangle$ hyperfine ground state, confined in a crossed optical dipole trap with harmonic trapping frequencies $2\pi\times (190,93,121)$ Hz along three orthogonal directions. The bias magnetic field is stabilized to 0.8 G with a feedback control loop, which gives $q_B=2\pi\times46$ Hz or $11.7|c_2|$. To initiate spin-nematic squeezing, $q$ is quenched to the value of $| c_2 |$ with the help of MW dressing. 
%The relative power stability of our MW source is about one thousandth, direct dressing under such a high bias magnetic field will therefore require a dressing MW induced QZS of $q_{\rm{MW}}=-10.7|c_2|$ with an accompanied fluctuation on the order 0.01$| c_2 |$, which is insufficiently stable and will compromise the resulting squeezing dynamics. 
The influence of  MW power fluctuation is mitigated by ramping the bias magnetic field down to 0.537 G in 300 ms at a reduced $q_B$ of $2\pi\times20.7$ Hz or $5.3|c_2|$ before initiating spin-nematic squeezing dynamics. The effective $q = q_{\rm{B}} + {q_{\rm{MW}}}$ coming from both the magnetic field and AC Stark shift of the dressing MW will accordingly have a noise of $0.004|c_2|$. To inhibit weak spin-mixing during the ramping process to smaller $q_{\rm{B}}$, we switch on a MW field 10.43 MHz red-detuned from the $| {1,0} \rangle$ to  $| {2,0} \rangle$  clock transition with about 6 W of power to keep $q$ above 10$\left| c_2 \right|$. Afterwards, the condensate is held for another 100 ms to ensure the magnetic field is stabilized. Ambient RF noise may transfer a tiny amount of atoms from  $| {1,0} \rangle$ to  $| {1,\pm 1} \rangle$ during the ramping. These atoms are removed by two resonant MW $\pi$ pulses coupling with the $| {1,\pm 1} \rangle$ to  $| {2,\pm 2} \rangle$ transitions, and then cleaned out with a flush of resonant probe light beam. The total atom number in the end is about $N=26400$ with a measured standard deviation of 240, which is 1.5 times that of the BEC shot noise of $\sqrt{N}$.

\subsubsection*{Standard quantum limit (SQL)}
Standard quantum limit (SQL) in two-mode interferometry is given by $1/\sqrt{N}$ for $N$ particles. For multi-mode, the precise value of a possible prefactor depending on the number of modes coupled in the probing process may exist. In the nonlinear Rabi interferometer implemented here, a RF pulse couples all of the three Zeeman states symmetrically and therefore realizes a three-mode SU(2) interferometer~\cite{Zou6381}, whose SQL is known to be $1/2\sqrt{N}$. Such an improved phase resolution by a factor of 2 with increased number of modes can be intuitively explained by the entanglement between electron spin $J=1/2$ and nuclear spin $I=3/2$ within a single particle, and is achieved with the Schr\"{o}dinger kitten state~\cite{Chalopin:2018aa} $\dfrac{1}{\sqrt{2}}\left(|m_J=\dfrac{1}{2}, m_I=\dfrac{1}{2}\rangle+|m_J=-\dfrac{1}{2}, m_I=-\dfrac{1}{2}\rangle\right)$, or equivalently $\dfrac{1}{\sqrt{2}}\left(|F=1,m_F=1\rangle+|F=1,m_F=-1\rangle\right)$ in the coupled angular momentum basis.\par
%Nevertheless, to have a fair comparison with previous experiments employing two levels, we reference the SQL in Fig.~2 to the two-mode SQL, and the reported metrological gain is consequently 6 dB higher with respect to the three-mode SQL.\par
For the nonlinear Ramsey interferometry employed here, phase encoding by $\hat{Q}_0$ operator leads to well determined enhancement factor or metrological gain. The encoding of quadrature phase can be realized by a two-level coupling, for example, between $|1,0\rangle$ and $|2,0\rangle$ transition\cite{Kruse}, essentially constituting of a two-mode interferometer. More strictly, the QCRB for $\hat{Q}_0$ and probe state $|\psi\rangle$ is formally given by $1/\sqrt{F_Q[\hat Q_0, |\psi\rangle]}$,
where  $F_Q=4(\Delta\hat Q_0)^2$ denotes the quantum Fisher information (QFI) upper bounded by $F_Q\leq N$ for a $N$-particle coherent spin state, and the SQL therefore becomes $1/\sqrt{N}$.

\subsubsection*{Spin rotation around $\hat Q_0$ axis}
Rotations around $\hat Q_0$ axis are employed to characterize the anisotropic distribution of spin-nematic squeezing and to realize effectively time-reversed dynamics and phase encoding in the nonlinear Ramsey interferometer. Its implementation is, however, not straight-forward at first sight, since we cannot directly synthesize the $\hat Q_0$ operator which involves coupling between $|1,\pm1\rangle$ spin components. We note nevertheless that such a spin rotation in the spin-nematic sphere is equivalent to related operation of $\hat N_0$ according to:
\begin{equation}
\begin{aligned}
&e^{-i\hat Q_0\theta}\hat L_xe^{i\hat Q_0\theta}=e^{-i\hat N_0\theta}\hat L_xe^{i\hat N_0\theta}=\hat L_x\cos\theta-\hat Q_{yz}\sin\theta,\\
&e^{-i\hat Q_0\theta}\hat Q_{yz}e^{i\hat Q_0\theta}=e^{-i\hat N_0\theta}\hat Q_{yz}e^{i\hat N_0\theta}=\hat Q_{yz}\cos\theta+\hat L_x\sin\theta,
\end{aligned}
\end{equation}
thus we can simply encode a phase on the $|1,0\rangle$ spin component and then realize the above rotations.\\
A variable phase can be imprinted to $|1,0\rangle$ state by two $\pi$ pulses resonantly coupled to $| {2,0} \rangle$ with an adjustable relative phase $\theta$. This approach is employed to rotate states around $\hat Q_0$ axis in both linear and nonlinear interferometries reported here (except for the quadrature phase interrogation in Ramsey interferometry experiment). The relative phase $\theta$ of the two MW pulses is directly mapped to the accumulated geometric phase of state $|1,0\rangle$ as $\pi-\theta$. Meanwhile, the MW pulses will also induce AC Stark shifts to all the three spin components due to the multi-level hyperfine level structure, thereby generating an extra change $\theta'$ to the actually obtained quadrature phase $\phi=(\phi_1+\phi_{-1})/2-\phi_0=\theta+\theta'$ where $\phi_i$ denotes the accumulated phase on $|1,i\rangle$. Fortunately, this extra phase is stable with fixed MW pulse areas, therefore can be considered as a constant offset. To calibrate the quadrature phase, we perform a linear Ramsey interferometry with polar state as the input, which is first transformed into the single particle state $\displaystyle | \psi  \rangle  =  -\frac{i}{2} | {1,-1} \rangle  + \frac{1}{{\sqrt 2 }} | {1,0} \rangle  - \frac{i}{2}| {1,1} \rangle $ after a $\pi/4$ pulse rotation around $\hat L_x$. Two resonant microwave $\pi$ pulses with equal Rabi frequency $2\pi\times25$ kHz then encode their relative phase on atoms, which corresponds to an atomic spin rotation around $\hat Q_{yz}$ axis after a second $\pi/4$ RF pulse around $-\hat L_x$ axis. The amplitudes of the two MW pulses follow the Blackman profile in order to reduce crosstalk among hyperfine spin components. The normalized $\langle\hat Q_0\rangle/N$, which reduces to $\langle\hat\rho_0\rangle-0.5$ if we take the arbitrary Larmor phase to be 0, is proportional to $\cos\phi$, from which one infers the offset of quadrature phase to be $\theta'=-0.03\pi$, based on the calibration data in Fig.~\ref{fig:calibration2}.
\begin{figure}[ht] 
	\centering
	\includegraphics[width=1\linewidth]{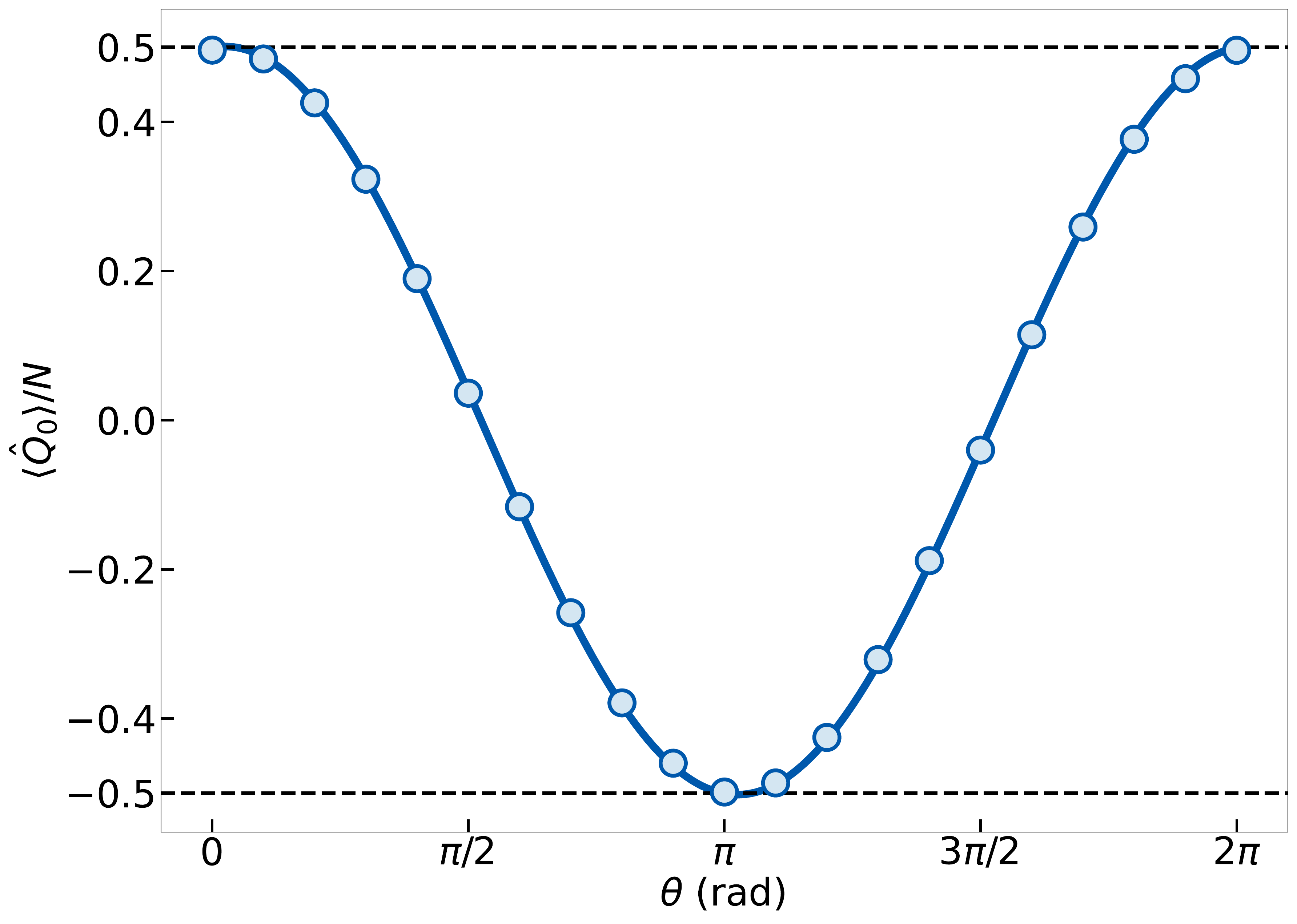}
	\caption{{\bf Calibration of spin rotation around $\hat Q_0$ axis.} Spin rotations are realized by adjusting the relative phase between two consecutive MW $\pi$ pulses, which are resonantly coupled to the clock transition between $|1,0\rangle$ and $|2,0\rangle$ hyperfine states. The blue dots are experimental results averaged over 10 repetitions and the uncertainty of one standard deviation is smaller than the dot size. The solid line is a fitted curve using trigonometry function which helps to infer the actually accumulated geometric phase from the MW pulses.}
	\label{fig:calibration2}
\end{figure}

\subsubsection*{Small-angle rotation around $\hat L_x$ axis}
RF pulses are utilized to rotate the probe state along $\hat L_x$ axis by a small angle $\phi$ in the nonlinear interferometer in Fig.~2. To precisely calibrate this rotation, we start with the coherent spin state $\exp(-i\hat L_x\pi/4)|0\rangle^{\otimes N}$ which is produced by applying a $\pi/4$ RF pulse to the polar state. This state is sensitive to small-angle Rabi rotation around $\hat L_x$ axis, the angle of which can be extracted from the change of $\langle\hat\rho_0\rangle$ through $\phi=\left[\arcsin(1-2{\rho _0^f}) - \arcsin(1-2{\rho _0^i})\right]/2$, where $\rho _0^{i(f)}\approx0.5$ corresponds to the mean fractional population of atoms in the $\left|0\right\rangle$ component before (after) spin rotation. We employ two consecutive resonant pulses with opposite rotation axes to encode a small angle, as in our previous work\cite{Zou6381}. The combination of the two pulses avoids the difficult-to-control switching effects over the required short switching time. Experimentally, we fix the amplitude of the first pulse while change the second one to adjust the net angle $\phi$. The experimental data is recorded in Fig.~\ref{fig:calibration}. The results obtained are not linearly dependent on the relative RF amplitude, possibly due to the saturation of the RF amplifier. For the experiments implemented in Fig.~2, we adopt these well-calibrated rotation angles to complete our interferometry measurements. 
\begin{figure}[ht] 
	\centering
	\includegraphics[width=1\linewidth]{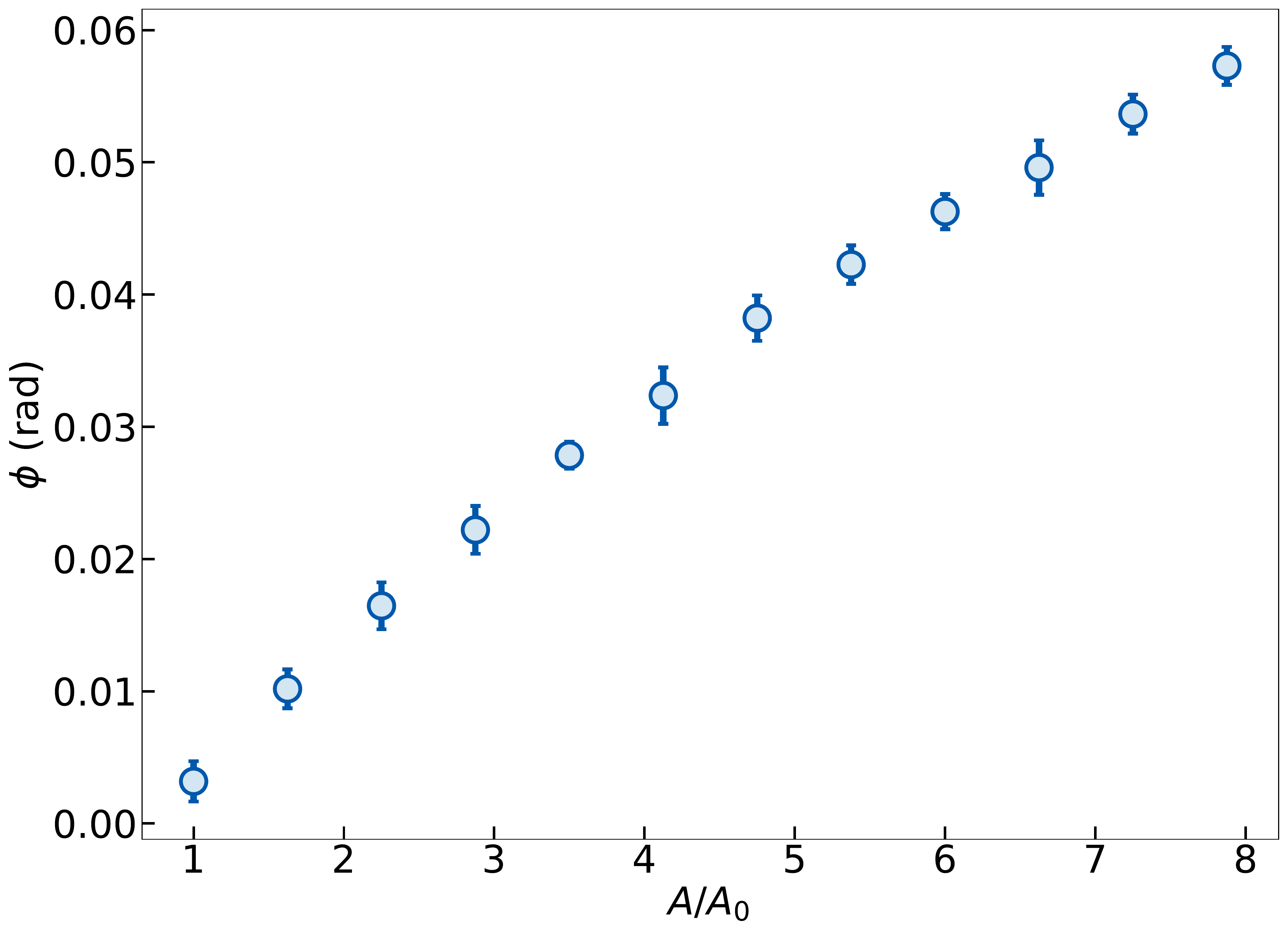}
	\caption{{\bf Calibration of small-angle Rabi rotation around $\hat L_x$ axis.}  In the experiment, we fix the RF amplitude ($A_0$) of the first in the composite pulse, while change the amplitude ($A$) of the second one to tune the net rotated angle. This figure shows the dependence of the accumulated phase angle and the amplitude ratio of $A/A_0$.}
	\label{fig:calibration}
\end{figure}

\subsubsection*{Phase interrogation in nonlinear Ramsey interferometer}
To precisely control the small quadrature phase as is required in the nonlinear Ramsey interferometer, we choose a different phase encoding approach which utilizes an off-resonant MW field to shift the energies of the three spin components\cite{Liu2021}. When the effective QZS is sufficiently large (compared to the spin-exchange strength $|c_2|$), the spin-mixing dynamics is suppressed and only a quadrature phase is interrogated. Like the small-angle Rabi rotation, here we also apply a composite pulse consisting of two single MW pulses. The two pulses are 3.1 MHz red and blue detuned from the 6.8 GHz clock transition between $|1,0\rangle$ and $|2,0\rangle$ respectively, giving rise to QZS of $\pm2\pi\times124$ Hz (or $\pm32|c_2|$). The amplitude of red-detuned pulse lasting for $10 \rm{\mu} s$ is optimized to cancel the phase shift induced by the rising and trailing edges of the blue-detuned one, therefore the net quadrature phase accumulated is $\phi=q\tau$, proportional to the duration of the blue-detuned MW pulse.

\subsubsection*{Atom number counting}
During the experiments, we routinely calibrate the absorption imaging system based on the method introduced in our earlier work\cite{Luo620}. To confirm the accuracy of atom number counting, we prepare BEC in the polar state, and then transfer the atoms to coherent superposition states of $|1\rangle$ and $\left|-1\right\rangle$ by applying a resonant $\pi/2$ RF pulse. Ideally, the fluctuation of atom number difference between $\left|\pm1\right\rangle$ modes should be $\Delta(N_1-N_{-1})=\sqrt{N}$ which is the quantum projection noise. However, imperfect imaging system may cause  a prefactor to the experimentally measured particle number, i.e., $N'=\alpha N$ with $\alpha\approx1$ to the first order of approximation, which therefore gives rise to a correction to projection noise scaling as $\Delta(N_1'-N_{-1}')=\sqrt{\alpha N'}$. We perform experimental measurements for $N$ ranging from $0$ to $32000$, as shown in Fig.~\ref{fig:sql}. The linear fit result gives $\alpha=0.98\pm0.05$, which therefore affirms the accuracy of our atom number counting.
%\clearpage
\begin{figure}[ht]
	\centering
	\includegraphics[width=1\linewidth]{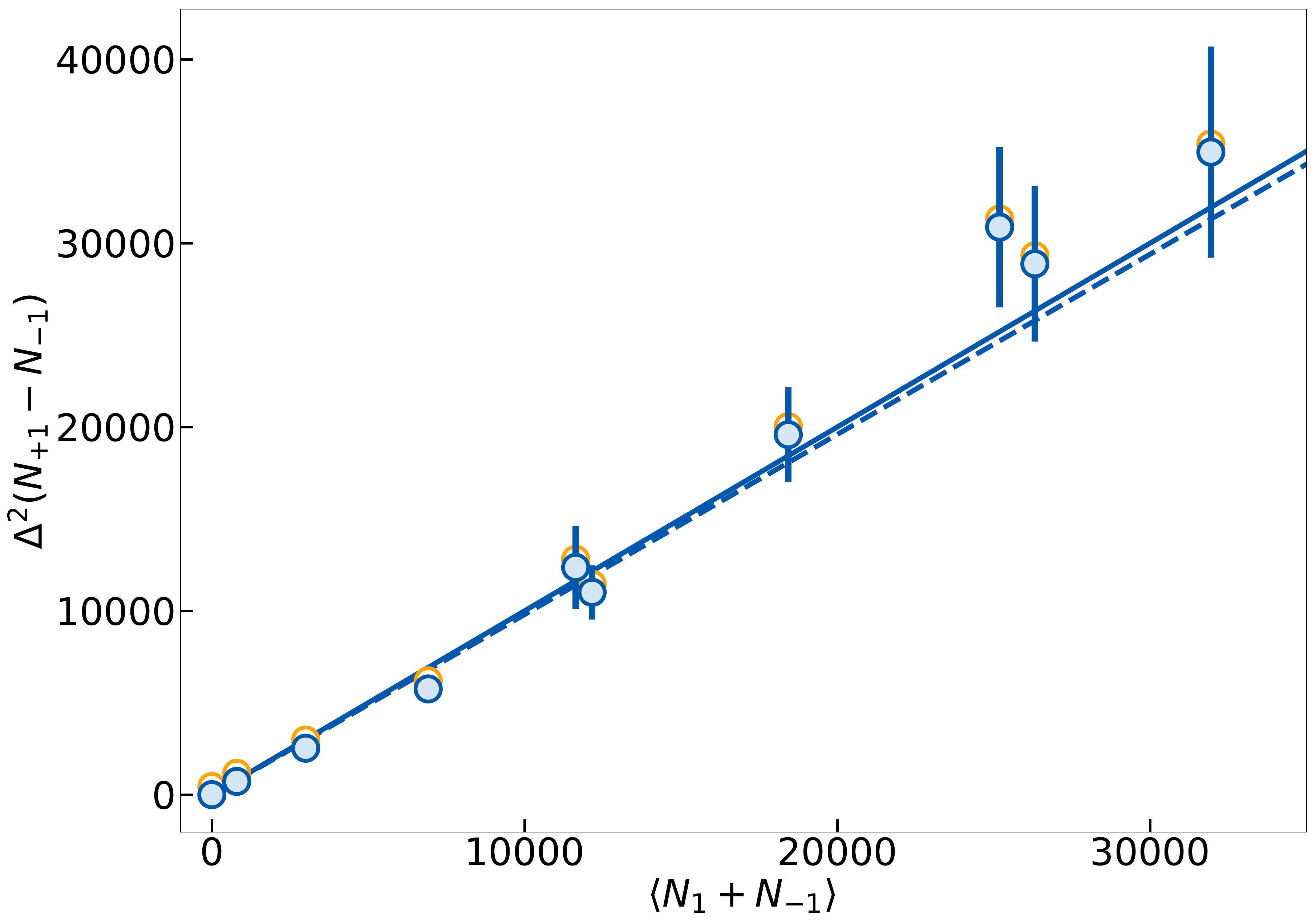}
	\caption{Calibration of atom number detection accuracy. The orange (blue) circles denote the measured standard deviation of the number difference between $|1,\pm1\rangle$ modes before (after) subtracting atom-independent detection noise of $\Delta\hat L_z^{\rm{DN}}=24$, obtained from 100 independent experiments respectively. The solid blue line represents the quantum projection noise of $\sqrt{N_1+N_{-1}}$, while the dashed line is the linear fitted curve.}
	\label{fig:sql}
\end{figure}

\end{document}

\title{Supplementary information: Quantum enhanced sensing by echoing spin-nematic squeezing in atomic Bose-Einstein condensate}
%	\author{Tian-Wei Mao$^{1,*}$, Qi Liu$^{1,*,\dagger}$, Xin-Wei Li$^1$, Jia-Hao Cao$^1$, Wen-Xin Xu$^1$, Meng Khoon Tey$^{1,2}$ , Yi-Xiao Huang$^{3,\ddagger}$ and Li You$^{1,2,4,\S}$}
	
%\affiliation{State Key Laboratory of Low Dimensional Quantum Physics,\\
%  Department of Physics, Tsinghua University, Beijing 100084, China}
%\affiliation{Frontier Science Center for Quantum Information, Beijing, China}
%\affiliation{School of Science, Zhejiang University of Science and Technology,\\
% Hangzhou, Zhejiang 310023, China}
%\affiliation{Beijing Academy of Quantum Information Sciences, Beijing 100193, China}
%$\dagger$ Present address: Laboratoire Kastler Brossel, Coll\`ege de France, CNRS, ENS-PSL University, Sorbonne Universit\'e, 11 Place Marcelin Berthelot, 75005 Paris, France.\par

	\author{Tian-Wei Mao$^*$}
	\affiliation{State Key Laboratory of Low Dimensional Quantum Physics, Department of Physics, Tsinghua University, Beijing 100084, China}
	\author{Qi Liu$^*$}
	\affiliation{Laboratoire Kastler Brossel, Coll\`ege de France, CNRS, ENS-PSL University, Sorbonne Universit\'e, 11 Place Marcelin Berthelot, 75005 Paris, France}
	\author{Xin-Wei Li}
	\affiliation{Graduate School of China Academy of Engineering Physics, Beijing 100193, China}
	\author{Jia-Hao Cao}
	\affiliation{State Key Laboratory of Low Dimensional Quantum Physics, Department of Physics, Tsinghua University, Beijing 100084, China}
	\author{Feng Chen}
	\affiliation{State Key Laboratory of Low Dimensional Quantum Physics, Department of Physics, Tsinghua University, Beijing 100084, China}
	\author{Wen-Xin Xu}
	\affiliation{State Key Laboratory of Low Dimensional Quantum Physics, Department of Physics, Tsinghua University, Beijing 100084, China}
	\author{Meng Khoon Tey}
	\affiliation{State Key Laboratory of Low Dimensional Quantum Physics, Department of Physics, Tsinghua University, Beijing 100084, China}
	\affiliation{Frontier Science Center for Quantum Information, Beijing, China}
	\affiliation{Collaborative Innovation Center of Quantum Matter, Beijing 100084, China}
	\affiliation{Hefei National Laboratory, Hefei, Anhui 230088, China}
	\author{Yi-Xiao Huang$^{\dagger}$}
    	\affiliation{School of Science, Zhejiang University of Science and Technology, Hangzhou, Zhejiang, 310023, China}
	\author{Li You$^{\ddagger}$}
	\affiliation{State Key Laboratory of Low Dimensional Quantum Physics, Department of Physics, Tsinghua University, Beijing 100084, China}
	\affiliation{Frontier Science Center for Quantum Information, Beijing, China}
	\affiliation{Collaborative Innovation Center of Quantum Matter, Beijing 100084, China}
	\affiliation{Hefei National Laboratory, Hefei, Anhui 230088, China}
	\affiliation{Beijing Academy of Quantum Information Sciences, Beijing 100193, China}

	\date{\today}
	\maketitle
	\section{Basics of spin mixing dynamics in spin-1 system}
	\subsection{SU(3) group description}\label{1A}
	The mean values of SU(2) group elements composed of spin angular momentum operators $\{\hat L_x, \hat L_y, \hat L_z\}$ provide complete descriptions for systems  of single two-level particles, which however, are insufficient to uniquely determine the quantum states for systems of three-level or spin-1 particles. For example, two orthogonal single-particle states $|0,1,0\rangle$ and $|1,0,1\rangle/\sqrt{2}$ of a spin-1 system can take the same vanished mean values, i.e., $\langle \hat L_i\rangle=0$ for all $i \in \{ x,y,z\}$, but are distinct. Additional observables, e.g., the so-called quadrupole or nematic tensor operators~\cite{Hamley2012,Iacopo,Yaomin} are therefore required to identify such systems. They are related to spin angular momentum operators through the nematic tensor operators
\begin{equation}
	{\hat Q_{ij}} = {\hat L_i}{\hat L_j} + {\hat L_j}{\hat L_i} - \frac{4}{3}{\delta _{ij}}\hat{\mathbbm{1}}_3,
\end{equation}
where $\delta _{ij}$ and $\mathbbm{1}_3$ are the Kronecker delta and unit operator respectively. A total of five linearly independent quadrupole operators exist, which take the matrix forms of:
\begin{widetext}
\begin{align}
	{\hat Q_{xz}} &=  \frac{1}{\sqrt2} \left( {\begin{array}{*{20}{c}}
		0&1&0\\
		1&0&{ - 1}\\
		0&{ - 1}&0
		\end{array}} 
		\right),
	{\hat Q_{yz}} =  \frac{1}{\sqrt2} \left( {\begin{array}{*{20}{c}}
		0&{-i}&0\\
		i&0&i\\
		0&{ - i}&0
		\end{array}} 
		\right),
	{\hat Q_{zz}} =  \frac{2}{3}\left( {\begin{array}{*{20}{c}}
		1&0&0\\
		0&{-2}&0\\
		0&0&1
		\end{array}} 
		\right)\\ 
	{\hat Q_{yy}} &= \left( {\begin{array}{*{20}{c}}
		-1/3&0&-1\\
		0&2/3&0\\
		-1&0&-1/3
		\end{array}} 
		\right),
	\hat Q_{xy} = \left( {\begin{array}{*{20}{c}}
		0&0&-i\\
		0&0&0\\
		i&0&0
		\end{array}} 
		\right).\notag
	\end{align}
\end{widetext}
Together with $\{\hat L_x,\hat L_y,\hat L_z\}$, they form the SU(3) group description of a spin-1 system, leading to a general density matrix for spin-1 uniquely parameterized by
\begin{equation}
	\hat \rho_0 = \frac{1}{3}\hat{\mathbbm{1}}_3 + \sum\limits_i l_i\hat{L}_i + \sum\limits_{ij} {{q_{ij}}{{\hat Q}_{ij}}}.
\end{equation}
To visualize the quantum state, we adopt the  spin-nematic Bloch sphere representation spanned by the spin angular momentum and nematic tensor operators $\{\hat Q_{yz}/2, \hat L_x/2, \hat Q_0=(\hat Q_{yy}-\hat Q_{zz})/4\}$. They follow SU(2) commutation relations explicitly as in the following
\begin{equation}
\begin{aligned}
&[\hat Q_{yz}/2,\hat L_x/2]=i\hat Q_0,\\
&[\hat L_x/2, \hat Q_0]=i\hat Q_{yz}/2,\\
&[\hat Q_0, \hat Q_{yz}/2]=i\hat L_x/2.
\end{aligned}
\end{equation}
This subspace provides a physically intuitive description for spin-mixing dynamics starting from the polar state of all atoms in $\left| F=1,m_F=0\right\rangle$~\cite{Hamley2012}. In the undepleted regime when $\langle\hat N_1+\hat N_{-1}\rangle\ll\langle\hat N_0\rangle$, the spin-mixing dynamics can be visualized by the changing quasi-probability distribution on spin-nematic sphere surface, which gives rise to an intuitive description for spin-nematic squeezing. Other single-particle operations involved in this study such as encoding of quadrature phase and radio-frequency (RF) Rabi oscillation can be simply illustrated as spin rotations around $\hat Q_0$ and $\hat L_x$ axis respectively. It is worth noting that the pre-factor of $1/2$ in front of $\hat L_x$ and $\hat Q_{yz}$ in the SU(2) subgroup maps a physical rotation of angle $\phi$ around $\hat L_x$-axis to a rotation of $2\phi$ on the spin-nematic sphere.
	
%\subsection{Spin-Nematic Squeezing and Effective Time-Reversal}\label{1B}
\subsection{Analytical solution for spin-nematic squeezing}\label{1B}
For condensed ${}^{87}$Rb atoms confined in a tight optical dipole trap, all the three spin components assume identical spatial wave function, determined by spin-independent interactions alone in the so-called single mode approximation (SMA) \cite{Law:1998aa}. The full spin-mixing dynamics is then described by the following Hamiltonian
	\begin{equation}
		\begin{aligned}
			{\hat H} = &\dfrac{c_2}{2N}\hat{\bf{L}}^2-q\hat{N}_0\\
			=&\dfrac{c_2}{2N}\big[2({\hat a}_1^{\dag}{\hat a}_{-1}^{\dag}{\hat a}_0{\hat a}_0+{\hat a}_1{\hat a}_{-1}{\hat a}_0^{\dagger}{\hat a}_0^{\dagger})+({\hat N}_1-{\hat N}_{-1})^2\\
			&+(2{\hat N}_0-1)({\hat N}_1+{\hat N}_{-1})\big]+q({\hat N}_1+{\hat N}_{-1}),
		\end{aligned}
		\label{ham}
	\end{equation}
where $c_2<0$ is the ferromagnetic spin-exchange interaction strength, and $q$ is the quadratic Zeeman shift (QZS). $\hat a_i(\hat a_i^{\dagger})$ and $\hat N_i$ stand for the annihilation~(creation) and number operator for the $i$-th ($i=0,\pm1$) spin component.
In the short-time limit for dynamics starting from polar state, atom number in $|0\rangle$ mode is undepleted, therefore $\hat a_0$ can be approximately  replaced by a $c$-number $\sqrt{N}e^{i\phi_0}$ leading to the approximate Hamiltonian
	\begin{equation}
		\hat H_{\rm udp} = c_2(e^{2i\phi_0}\hat a_{1}^\dagger \hat a_{-1}^\dagger +h.c.) + (q+c_2)(\hat N_{1} +\hat N_{-1}).
		 \label{SU11ham}
	\end{equation}
%	It is interesting to note that this reduced Hamiltonian can be easily sign-flipped by adding a phase of $\pi/2$ to $\phi_0$ together with adjusting $q$ to $-2c_2-q$ without needing to actually reversing the sign of $c_2$, as has been demonstrated in a previous work~\cite{Linnemann:2016aa}. 
	To describe the build-up of spin-nematic squeezing, we set $\phi_0=0$ without loss of generality, and define $\mathcal{\hat H} =\hat H_{\rm udp}/|c_2|$ as well as $\delta = q/|c_2| -1$. Spin-mixing dynamics then takes place for $\delta\in[-1,1]$, i.e., for  $q\in [0,2\left| {{c_2}} \right|]$. In the Heisenberg picture, the equations of motion for $\hat a_{\pm1}$ operators become
	\begin{equation}
		\begin{aligned}
		i{\dot {\hat a}_{ + 1}} &= [{\hat a_{ + 1}},\mathcal{\hat H}] = \hat a_{-1}^\dagger + \delta \hat a_{+1},\\
		i{\dot {\hat a}_{ - 1}} &= [{\hat a_{ - 1}},\mathcal{\hat H}] = \hat a_{+1}^\dagger + \delta \hat a_{-1},
		\end{aligned}
	\end{equation}
	with the following analytical solutions
%	\begin{widetext}
	\begin{equation}
		\begin{aligned}
		\hat a_{+1}(t) &= [\cosh (kt) -\frac{\delta}{k}\sinh (kt)] \hat a_{+1} (0) + \frac{i}{k} \sinh (kt) \hat a_{-1}^\dagger (0), \\
		\hat a_{-1}(t) &= [\cosh (kt) -\frac{\delta}{k}\sinh (kt)] \hat a_{-1} (0) + \frac{i}{k} \sinh (kt) \hat a_{+1}^\dagger (0), 
		\end{aligned}
		\label{HSFirst}
	\end{equation}
%\end{widetext}
where $k = \sqrt{1 - \delta ^2}$. Meanwhile, $\hat L_x$ and $\hat Q_{yz}$ reduce to their corresponding two-mode operators:
\begin{equation}
	\begin{aligned}
	 \hat L_x &= \sqrt{\frac{N}{2}}[(\hat a_{+1}^\dagger + \hat a_{-1}^\dagger) + (\hat a_{+1} + \hat a_{-1})],\\
	 \hat Q_{yz} &= \sqrt{\frac{N}{2}}[-i(\hat a_{+1}^\dagger + \hat a_{-1}^\dagger) + i(\hat a_{+1} + \hat a_{-1})].
	\end{aligned}
\end{equation}
For a clear manifestation of spin-nematic squeezing, we define the quadrature operator in the $\hat L_x$-$\hat Q_{yz}$ plane:
\begin{equation}
\hat Q(\theta) =\sin\theta\hat Q_{yz}- \cos\theta\hat L_x.
\end{equation}
It is then straightforward to see that the spin angular momentum operator $\hat L_x=\hat Q(\pi)$ and nematic tensor operator $\hat Q_{yz}=\hat Q(\pi/2)$ mutually convert into each other upon a change of quadrature phase $\pi/2$. The variance of $\hat Q$ is found to be based on Eq. (\ref{HSFirst}) 
\begin{equation}
	\begin{aligned}
	\Delta ^2\hat Q{(\theta)}/N &= \cosh ^2(kt) +(1 + \delta^2)\frac{\sinh ^2(kt)}{k^2} \\ 
	&+ \left[\frac{\sinh (2kt)}{k}\sin (2\theta) + \frac{2\delta\sinh^2(kt)}{k^2}\right],
	\end{aligned}
	\label{quradrapolesqueezing}
\end{equation}
which arrives its minimum and maximum respectively at
\begin{equation}
\begin{aligned}
	&\theta_{\rm{min}} = \dfrac{3\pi}{4} - \dfrac{1}{2}\arctan\left[\dfrac{\delta}{k}\tanh(kt)\right],\\
	&\theta_{\rm{max}} = \dfrac{\pi}{4} - \dfrac{1}{2}\arctan\left[\dfrac{\delta}{k}\tanh(kt)\right].
	\label{optimal}
	\end{aligned}
\end{equation}
Specifically, for $q = \left| c_2 \right|$ (or $\delta=0$) as explored in our experiment, Equation (\ref{optimal}) reduces to time-independent values of $\theta_{\rm{min}} =3\pi/4$ and $\theta_{\rm{max}}=\pi/4$. The minimal and maximal variance of $\hat Q$ become
 \begin{equation}
 \begin{aligned}
	&\Delta ^2Q_{\rm{min}}/N = \cosh(2t) -\sinh(2t)\leq1,\\
	&\Delta ^2Q_{\rm{max}}/N = \cosh(2t) +\sinh(2t)\geq1,
\end{aligned}
 \end{equation}
 For polar state $|0,N,0\rangle$ at $t=0$, this fluctuation distribution is isotropic, or $\Delta^2Q/N=1$ for all $\theta$. A quadrature fluctuation of less than 1 confirms that squeezing is generated during spin-mixing dynamics.

\subsection{Numerical simulations of spin-mixing dynamics in the presence of noise}
%To simulate the spin-mixing dynamics in the ideal case, one can directly diagonalize the Hamiltonian in the Fock basis and numerically solve the Schr\"{o}dinger equation. 
The above analytical results describe the essential physics for spin-nematic squeezing in the restricted regime. To enable quantitative comparisons with experimental results, it is necessary to take into account noises associated with of atom loss and quadratic Zeeman shift (QZS) in our experiments. In this work, we perform simulations to solve for the stochastic differential equations (SDEs) resulting from the semi-classical method based on truncated Wigner approximation (TWA). Such an approach is found to work well in simulating the long-term spin-mixing dynamics previously~\cite{Liu2021}. A brief introduction to the main ideas of this approach is provided in the following.\par
By mapping the field operators $\hat a_i$ and $\hat a_i^\dagger$ into complex number $\psi_i$ and $\psi_i^*(i=0,\pm1)$ and adopting TWA, the quantum master equation including  atom loss and fluctuation of QZS is transformed into SDEs taking the following form
	\begin{widetext}
		\begin{equation}
			\left\{
			\begin{aligned}\label{rfsde}
				d\psi_1 &=-ic_2'\left[\psi_0^2\psi_{-1}^*+(|\psi_1|^2-|\psi_{-1}|^2+|\psi_0|^2)\psi_1\right]dt-\frac{\gamma}{2}\psi_1dt+\sqrt{\frac{\gamma}{2}}d\xi_1(t), \\
				d\psi_0 &=-ic_2'\left[2\psi_1\psi_{-1}\psi_0^*+\left(|\psi_1|^2+|\psi_{-1}|^2\right)\psi_0\right]dt+i\psi_0\left[qdt+\eta_qd\xi_q(t)\right]-\frac{\gamma}{2}\psi_0dt+\sqrt{\frac{\gamma}{2}}d\xi_0(t),\\
				d\psi_{-1} &=-ic_2'\left[\psi_0^2\psi_1^*+(|\psi_{-1}|^2-|\psi_1|^2+|\psi_0|^2)\psi_{-1}\right]dt-\frac{\gamma}{2}\psi_{-1}dt+\sqrt{\frac{\gamma}{2}}d\xi_{-1}(t), 
			\end{aligned}
			\right.
		\end{equation}
	\end{widetext}
where $c_2^\prime={c_2(t)}/{N(t)}$ denotes the drifting spin-exchange rate due to atom loss. $\gamma$ and $\eta_q$ denote the loss rate and noise strength of QZS respectively. $d\xi_i(t)$ and $d\xi_q(t)$ are complex Wiener noise increments satisfying $\overline{d\xi_i(t)}=\overline{d\xi_q(t)}=0$, $\overline{d\xi_i^*(t)d\xi_j(t)}=\delta_{i,j}dt$, and $\overline{d\xi_q^*(t)d\xi_q(t)}=dt$. The SDEs provide a stochastic approach for simulating the interacting spin system with classical noise, they are capable of  describing evolutions after averaging over repeated sampling. The effect of quantum noise is incorporated into the probability distribution of the initial state, which is given by the Wigner function and can be sampled according to
\begin{equation}
\begin{aligned}
&\psi_{\pm1}=\dfrac{1}{2}(\alpha_{\pm1}+i\beta_{\pm1}),\\
&\psi_0=\sqrt{N}+\dfrac{1}{2}(\alpha_{\pm0}+i\beta_{\pm0}),
\end{aligned}
\end{equation}
for polar state with $N$ atoms, where $\alpha_i$ and $\beta_i$ are independent real numbers following standard normal distribution. We typically take 10000 samplings in our simulations, each of which provides an evolution trajectory. 
The average and variance of mode population are then obtained from the average of simulated  trajectories, calculated according to $\langle{\hat N}_i\rangle=\overline{\psi_i^*\psi_i}-1/2$ and $(\Delta \hat {N}_i)^2=(\Delta\psi_i^*\psi_i)^2-1/4$.\par

The parameters used in the simulations include initial atom number $N=26400$, $c_2=-2\pi\times3.9$ Hz, $q=2\pi\times3.8$ Hz, loss rate $\gamma=0.069~\rm{s^{-1}}$ and noise strength of QZS $\eta_q=2\pi\times0.015~\sqrt{\rm{mHz}}$, all of which are independently calibrated in experiments. We also experimentally compensate for the influence of atom loss by ramping $q$ with a rate $\gamma_q=\gamma_c$ to keep $c_2(t)/q(t)$ fixed as in Ref.~\cite{Liu2021}.\par

%\subsection{Effective Time-Reversal and Quantum Phase Magnification}
\section{Spin-nematic nonlinear interferometry}
\subsection{Realization of effective time-reversal}
Nonlinear interferometry typically calls for effectively time-reversed dynamics to disentangle the entangled probe state in order to  achieve signal amplification. In spinor BEC, such an operation in its simplest form requires a sign-flip of spin-exchange strength $c_2$, which is unfortunately infeasible in the current system. However, in the undepleted pump regime where $\langle\hat N_0\rangle\approx N$, the effective time-reversal can be approximately realized by state rotation together with a quench of QZS. We can easily verify this from Eq. (\ref{SU11ham}), the sign of which can be flipped by encoding a phase of $\pi/2$ on $|0\rangle$ mode together with a quench of QZS from $q$ to  $-2c_2-q$. Specifically, for $q=-c_2$ as adopted in our experiments, the quenching process is no longer required and the time-reversal operation arises with a $\pi/2$ phase encoding alone or state rotation of $\pi/2$ around $\hat Q_0$ axis, as first demonstrated in the atomic SU(1,1) interferometry~\cite{Linnemann:2016aa}.
%To clearly see this, we rewrite the Hamiltonian in Eq.(\ref{ham}) as:
%\begin{equation}
%\begin{aligned}
%{\hat H} = &\dfrac{c_2}{N}({\hat a}_1^{\dag}{\hat a}_{-1}^{\dag}{\hat a}_0{\hat a}_0+{\hat a}_1{\hat a}_{-1}{\hat a}_0^{\dagger}{\hat a}_0^{\dagger})\\
%+&\left[\dfrac{c_2}{2N}(2{\hat N}_0-1)+q\right]({\hat N}_1+{\hat N}_{-1}),
%\end{aligned}
%\end{equation}
%The sign of the first term describing spin-mixing dynamics can be exactly flipped by encoding a phase of $\pi/2$ on $|0\rangle$ mode, which transforms $\hat a_0$ to $\hat a_0e^{i\pi/2}$. The second term reduces to $c_2+q$ in the undepleted regime, and can also be sign-flipped by quenching $q$ to $-2c_2-q$. Specifically, for $q=-c_2$ as adopted in our experiments, the quenching process is vanished and the time-reversal operation is achieved with a simple phase encoding or spin rotation of $\pi/2$ around $\hat N_0$ (or $\hat Q_0$) axis. 

\subsection{Quantum Magnification and Phase Sensitivity}
Time-reversed nonlinear dynamics can be utilized to amplify signal in nonlinear interferometry. To clearly appreciate this, we consider the protocol discussed in Fig.~2 of the main text, which involves a rotation of $\phi$ around $\hat L_x$ axis sandwiched in between the two state rotations of $\pi/4$ around $\hat Q_0$ axis. For simplicity, we take the balanced interferometric configuration in which the durations of nonlinear splitting and nonlinear recombining are the same and denoted by $t$. For $q=|c_2|$ as considered in the following, according to Eq.~(\ref{HSFirst}), $\hat a_{+1} (t)$ and $\hat a_{-1}(t)$  after the nonlinear splitting operation takes the forms
	 \begin{equation}
		 \begin{aligned}
			\hat a_{ + 1}(t) &= \cosh (t)\hat a_{+1}(0) + i\sinh (t) \hat a_{- 1}^\dagger(0),\\
			\hat a_{ - 1}(t) &= \cosh (t)\hat a_{-1}(0) + i\sinh (t) \hat a_{+ 1}^\dagger(0).
		 \end{aligned}
	 \end{equation}
The first quadrature phase encoding increases the phase of $|0\rangle$ mode by $\pi/4$, which gives
\begin{equation}
	 \hat L_x = \sqrt{\dfrac{N}{2}}\left[e^{i\pi/4}(\hat a_{+1}^\dagger + \hat a_{-1}^\dagger) + \rm{h.c.}\right],
\end{equation}
after $\hat a_0$ is replaced by $\sqrt{N}e^{i\pi/4}$. The following rotation of angle $\phi$ around $\hat L_x$ shifts the annihilation operators according to
\begin{equation}
	\begin{aligned}
		\hat a_{ + 1}(t + \tau ) &= \hat a_{ + 1}(t ) - i\sqrt{\frac{N}{2}}e^{i\pi/4}\phi,\\
		\hat a_{ - 1}(t + \tau ) &= \hat a_{ - 1}(t ) - i\sqrt{\frac{N}{2}}e^{i\pi/4}\phi,\\
	\end{aligned}
\end{equation}
where $\tau$ denotes the duration of rotation or the interrogation time and scales with $\phi$. Finally, the second $\hat Q_0(\pi/4)$-rotation effectively flips the sign of Hamiltonian, the final expression can hence be obtained again through Eq.~(\ref{HSFirst}) as
\begin{equation}
	\begin{aligned}
		\hat a_{ + 1}(2t + \tau ) &= \cosh (t) \hat a_{ + 1}(t + \tau) - i\sinh (t) \hat a_{ - 1}^\dagger( t + \tau),\\
		\hat a_{ - 1}(2t + \tau ) &= \cosh (t) \hat a_{ - 1}(t + \tau) - i\sinh (t) \hat a_{ + 1}^\dagger(t + \tau).
	\end{aligned}
	\label{nematicanalytic}
\end{equation}
The rotation angle $\phi$ in our nonlinear interferometer can be inferred from measuring the mean values of $\hat L_x$ and $\hat Q_{yz}$, which are related to the $\phi$ through
\begin{equation}
\begin{aligned}
&\langle\hat L_x\rangle=-\sqrt{2}N\phi\left[\cosh(t)+\sinh(t)\right],\\
&\langle \hat Q_{yz}\rangle=\sqrt{2}N\phi\left[\cosh(t)+\sinh(t)\right].\\
\end{aligned}
\end{equation}
	\begin{figure}[t]
		\centering
		\includegraphics[width=1\linewidth]{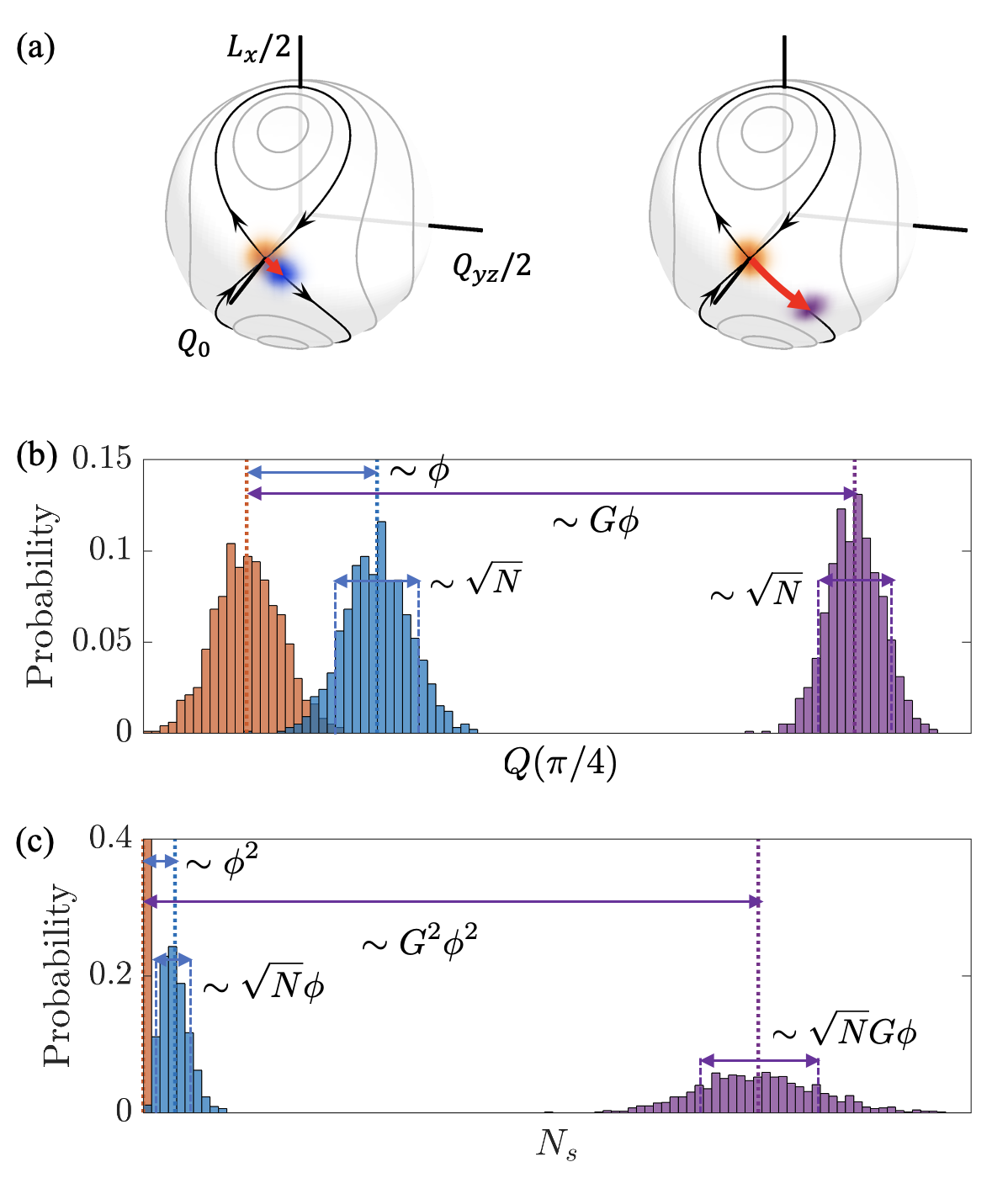}
		\caption{Quantum magnification in nonlinear interferometry. (a) Left: Initial polar state is displaced along $\hat Q(\pi/4)$ axis by spin rotation in linear interferometry. Right: This displacement can be magnified in nonlinear interferometry based on effectively time-reversed spin-nematic squeezing. Orange, blue and purple shade illustrate the initial state, final states without or with magnification respectively. (b) Transformation of the probability distribution by $Q(\pi/4)$. The mean value is amplified by a factor of $G$ while the noise remains about the same, leading to SNR enhanced $G$. (c) Transformation of the probability distribution of $N_s$. The amplification factor of the mean value is $G^2$, which is $G$ times larger than that of noise, giving rise to the same SNR enhancement. Dotted lines in (b) and (c) mark the mean values of the corresponding operators.}
		\label{mag}
	\end{figure}
In the absence of nonlinear splitting and recombining operations (namely for $t=0$), $\langle\hat L_x\rangle=-\langle\hat Q_{yz}\rangle=-\sqrt{2}N\phi$ can be treated as a reference signal from classical linear interferometry. In contrast, the time-reversed spin-nematic squeezing magnifies the $\phi$-dependent shift, which is magnified along $\hat Q(\pi/4)$ axis by a factor of $G=\cosh(t)+\sinh(t)$. On the other hand, the fluctuation of quadrature operator $\hat Q(\theta)$ remains at $\sqrt{N}$, independent of the quadrature angle $\theta$. According to the error propagation formula, the finally achieved phase sensitivity becomes
\begin{equation}
(\Delta\phi)_{Q(\pi/4)}=\dfrac{\Delta\hat Q(\pi/4)}{|d\langle \hat Q(\pi/4)\rangle/d\phi|}=\dfrac{1}{2\sqrt{N}G},
\end{equation}
which improves by a factor of $G$ the three-mode SQL of $1/2\sqrt{N}$ for rotation sensing by the polar state (see Fig.~\ref{mag}(b)).\par
In addition to the shift of $\langle\hat Q(\pi/4)\rangle$ in the spin-nematic plane, change of the total particle number in $|\pm1\rangle$ modes ($N_s=N_1+N_{-1}$) can also be employed to infer the encoded phase (see Fig.~\ref{mag}(c)). Based on the above discussion, we find
\begin{equation}{}
\langle N_s\rangle=\Delta^2N_s=NG^2\phi^2,
\end{equation}
which leads to a quantum enhanced phase sensitivity 
\begin{equation}
(\Delta\phi)_{N_s}=\dfrac{\Delta\hat N_s}{|d\langle\hat N_s\rangle/d\phi|}=\dfrac{1}{2\sqrt{N}G}.
\end{equation}
Although the two observables provide the same phase sensing precision, their specific mechanisms for quantum phase magnification differ slightly. Based on  the measurement of $\hat Q(\pi/4)$, the signal gets magnified by a factor of $G$ without inadvertently increasing quantum noise. While for the measurement of $\hat N_s$, both the signal and noise are amplified, but the magnification factor of signal ($\langle\hat N_s\rangle/\langle\hat N_s(t=0)\rangle=G^2$) is $G$ times larger than that of noise ($\Delta\hat N_s/\Delta\hat N_s(t=0)\rangle=G$), hence the signal-to-noise ratio (SNR) is $G$ times enhanced, while the enhancement of phase sensitivity remains the same. Finally, we want to emphasize that although  the two observables give the same SNR under ideal conditions, their comparisons become different when the experimental noises are taken into considerations. Measurement results of $\langle\hat N_s\rangle$ (or equivalently, $\hat\rho_0$) exhibit strong robustness to first-order magnetic field fluctuations during the nonlinear recombining as it commutes with $\hat{L}_z$, which is therefore chosen as the observables in our experiments.\par

	\begin{figure*}[t]
		\centering
		\includegraphics[width=1\linewidth]{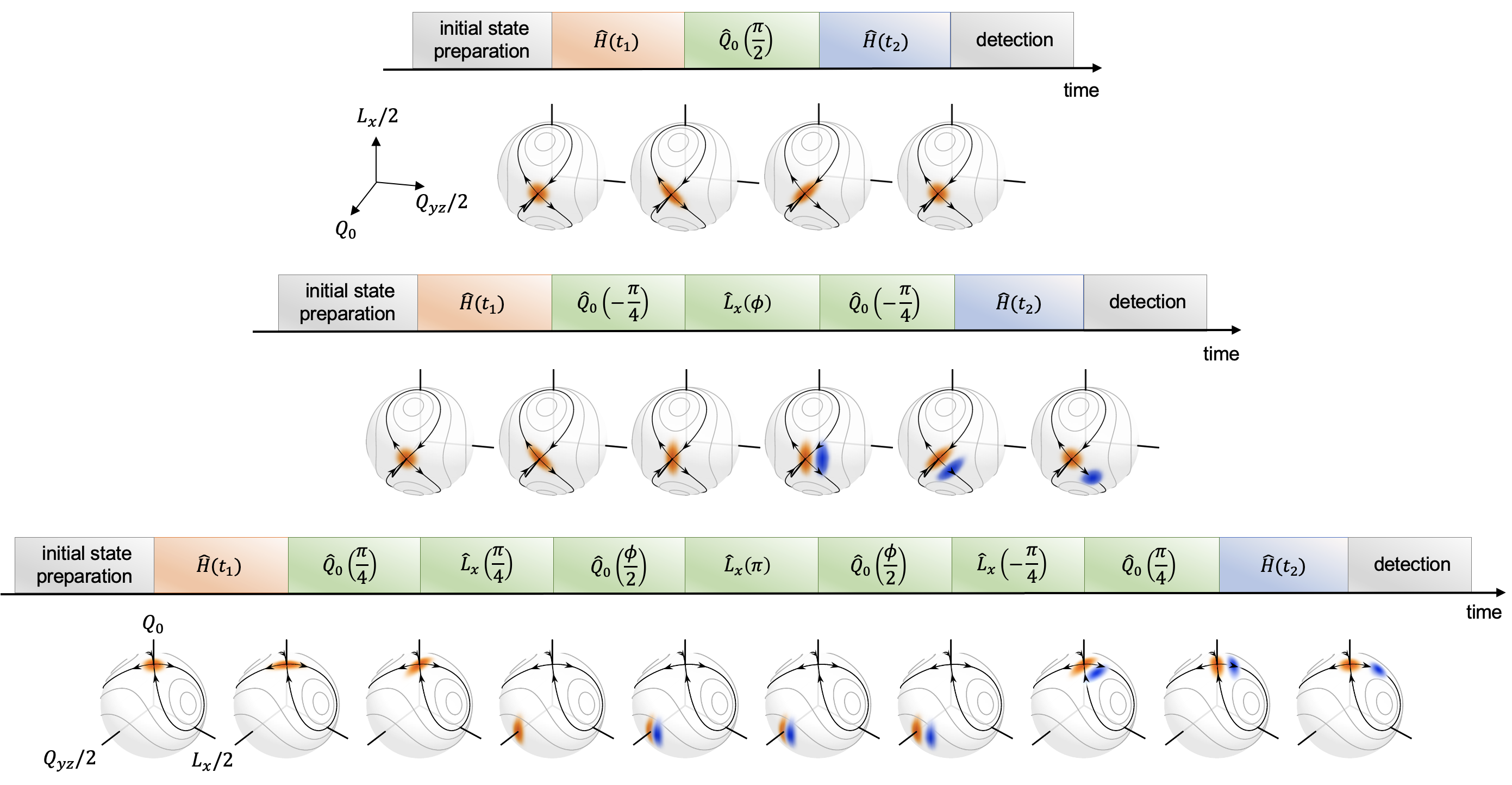}
		\caption{Operation sequences for the discussed nonlinear interferometers. Upper panel: the basic protocol without phase encoding.  Middle panel: nonlinear Rabi interferometer implemented by nesting the $L_x$-rotation in between $\hat{Q}_0(\pi/2)$. Lower panel: Nonlinear Ramsey interferometer for sensing the quadrature phase. }
		\label{settings}
	\end{figure*}
	
\subsection{Frameworks of nonlinear interferometry}
We present a concise framework for the implemented nonlinear interferometers in Fig.~\ref{settings}. Starting from the basic configuration with time-forward and effectively time-reversed spin-nematic squeezing (upper panel), we first implement a Rabi-type rotation sensing interferometer by nesting a rotation around $\hat L_x$ axis in between the nonlinear splitting and nonlinear recombining operations (middle panel). This rotation is then expanded into a Ramsey process (shown in lower panel) for sensing the quadrature phase, in which a pair of $\pm\pi/4$-pulses around $\hat L_x$ axis rotate the spin-nematic squeezed state to the equator of Bloch sphere for phase encoding and then back to the north pole for nonlinear readout.

\subsection{Comparisons with previous nonlinear interferometries}
	\begin{figure*}[t]
		\centering
		\includegraphics[width=1\linewidth]{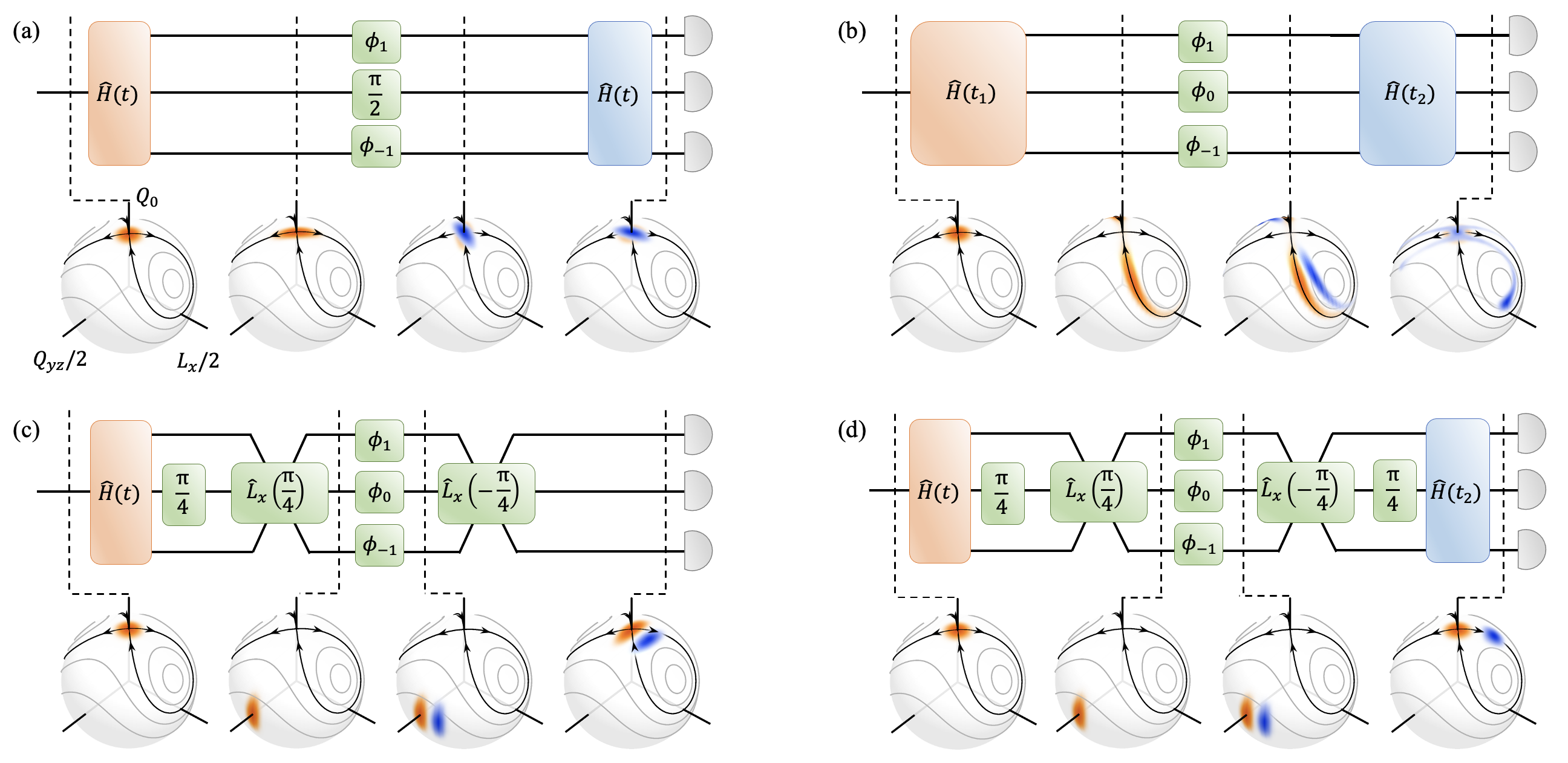}
		\caption{Comparisons with other entanglement enhanced interferometers implemented in spin-1 BEC. (a) SU(1,1) nonlinear atomic interferometry. Spin-nematic squeezing (or parametric amplification) generates a squeezed state (or two-mode squeezed vacuum state) for sensing the quadrature phase. The squeezing dynamics can be effectively time-reversed by encoding a phase of $\pi/2$ on the pump mode. (b) Cyclic nonlinear interferometry. The configuration is similar to SU(1,1) interferometry, except that the effective time-reversal is approximately achieved by the quasi-cyclic dynamics of the interacting  spin system. (c) Linear interferometer with spin-nematic squeezed state. The combination of a $\pi/4$-phase imprinting on pump mode and a RF $\pi/4$ pulse couples the three spin components before phase sensing, aligning the anti-squeezing direction of probe state parallel to $\hat{Q}_0$, therefore making the system highly sensitive to quadrature phase accumulation. (d) Spin-nematic nonlinear interferometer. Effectively time-reversed spin-nematic squeezing is adopted here to enhance the robustness to detection noise. The blue and orange regions on the Bloch spheres stand for the state quasi-probability distributions with or without phase encoding respectively.}
		\label{compare}
	\end{figure*}
In this subsection, we provide a conceptual and qualitative comparison of the spin-nematic nonlinear Ramsey interferometry with several closely related linear or nonlinear interferometers implemented in atomic BEC, including (a) nonlinear atomic SU(1,1) interferometry~\cite{Linnemann:2016aa}, (b) nonlinear interferometry based on cyclic dynamics~\cite{Liu2021}, and (c) linear interferometry with squeezed vacuum input state~\cite{Kruse}.
% (d) nonlinear interferometry based on sign-flipped Hamiltonian~\cite{Colombo:2022tz}. 
These configurations as well as their state evolutions are illustrated on the spin-nematic Bloch sphere in Fig.~\ref{compare}.\par
(a) Nonlinear atomic SU(1,1) interferometry. Atomic SU(1,1) interferometry is the first with both nonlinear splitter and recombiner implemented in a BEC system~\cite{Linnemann:2016aa}. The basic building blocks include a pair of spin-nematic squeezing (also known as parametric amplification) with a phase encoding operation sandwiched in-between. In the undepleted pump regime where $\langle\hat{N}_1+\hat N_{-1}\rangle\ll N$, the SU(1,1) Hamiltonian $\hat{H}_{\rm{SU(1,1)}}\sim e^{i2\phi_0}(\hat{a}_1^{\dagger}\hat{a}_{-1}^{\dagger}+\hat{a}_1\hat{a}_{-1})$ can be sign-flipped from changing the phase $\phi_0$ encoded on the pump mode by $\pi/2$ to realize effectively time-reversed spin-nematic squeezing. The accumulated quadrature phase $(\phi_1+\phi_{-1})/2$ is finally mapped to the total population in $|\!\pm\!1\rangle$ components, and can be inferred with a Heisenberg limited precision $\sim1/\langle\hat{N}_1+\hat{N}_{-1}\rangle$. However, one can notice that in the undepleted regime, even this Heisenberg limited bound is typically not much than the classical limit with respect to the total number of atoms, as the majority of atoms in the pump mode do not contribute to phase sensing in this case. We can also understand this feature on the Bloch sphere as shown in Fig.~\ref{compare}(a). During the phase encoding operation, the state stays around the north pole, therefore the quantum Fisher information, which is proportional to the variance of $\hat Q_0$, is restricted to a rather small value.\par
(b) Nonlinear cyclic interferometry. The low absolute phase sensitivity in atomic SU(1,1) interferometry discussed above can be improved with the nonlinear cyclic interferometry \cite{Liu2021}. In this configuration, the effective time-reversal is approximately realized by employing quasi-periodic spin-mixing dynamics, which brings the entangled state from spin-nematic squeezing  back to the vicinity of initial one after one quasi-period. Since this allows nonlinear splitting to work beyond the undepleted pump regime, the cyclic nonlinear interferometer can employ deeply entangled non-Gaussian state for phase sensing and beat the SQL with respect to the total number of atoms (see Fig.~\ref{compare}(b)). The main disadvantage of nonlinear cyclic interferometry lies on the fact the achieved metrological gain is highly sensitive to dynamical noises such as particle loss or fluctuation of QZS, as a result of the long evolution time required for nonlinear splitting and recombing.\par
(c) Linear interferometer with spin-nematic squeezed state. Spin-nematic squeezed state (or two-mode squeezed vacuum state) has been employed to improve the precision of a prototype atomic clock beyond SQL~\cite{Kruse}. Different from SU(1,1) interferometry, the generated squeezed state is first rotated around $\hat Q_0$ axis by $\pi/4$ to align the anti-squeezing direction parallel to $\hat L_x$ axis. A followup $\hat{L}_x(\pi/4)$-pulse then couples the three spin components and rotates the state to the equator of spin-nematic Bloch sphere (see Fig.~\ref{compare}(c)). The resulting state is squeezed along $\hat Q_{yz}$ axis while anti-squeezed along $\hat{Q}_0$ axis, hence is highly sensitive to quadrature phase encoded by microwave (MW) pulses transferring atoms between the $|1,0\rangle$ and $|2,0\rangle$ clock states. The interferometry sequence ends with a $-\pi/4$ rotation around $\hat L_x$ axis, which maps the encoded phase to the fractional population in $|0\rangle$ mode. Such a sequence is very similar to our spin-nematic nonlinear interferometry protocol shown in Fig.~\ref{compare}(d), except that there is no effectively time-reversed dynamics for nonlinear recombining that achieves signal magnification in the linear interferometry discussed. Therefore, the precision obtained is sensitive to detection noise, with limited metrological gains below 7 dB reported in earlier experiments~\cite{Kruse}.\par
%(d) Nonlinear interferometry based on sign-flipped Hamiltonian. Effective time-reversal can be exactly implemented by flipping the sign of Hamiltonian. Compared to the rotation based effective time-reversal, there is no principle limit of evolution time with this method, which therefore can be used to disentangle non-Gaussian entangled states as demonstrated in recent experiments. However, it is usually challenging to flip the sign of interaction strength in a many-body system, which poses a practical limitation to the application of this method. In addition, entangled non-Gaussian state is relatively difficult to prepare and detection, and also more fragile to environmental noises such as decoherence and dissipation, compared to squeezed state. We numerically find in the squeezed regime, both the methods give the same metrological gain. Therefore, our protocol provides an economic and efficient way to make the full exploitation of squeezed states for quantum metrology.\par

%The spin-nematic interferometer combines the advantages of the above two nonlinear interferometries, i.e., short duration of SU(1,1) and high QFI of cyclic. This is achieved by a linear coupling of the three spin components after the spin-nematic squeezing, as originally proposed in Ref. The linear coupling, or spin rotation aligns the anti-squeezing direction with $\hat Q_0$, therefore makes the probe state highly sensitive to $\hat Q_0$ rotation or quadrature phase encoding. 
\section{Numerical Analysis}
%We present some detailed numerical analysis in this section, including the dependance of metrological gain on experimental parameters ($t_1$, $t_2$ and $q$), robustness to detection noise, selection of observable operator, comparison of linear and nonlinear readout protocols and the influence of technical noise on the time-point of effective time-reversal.
\subsection{Robustness to detection noise}
\begin{figure}[t]
	\centering
	\includegraphics[width=1\linewidth]{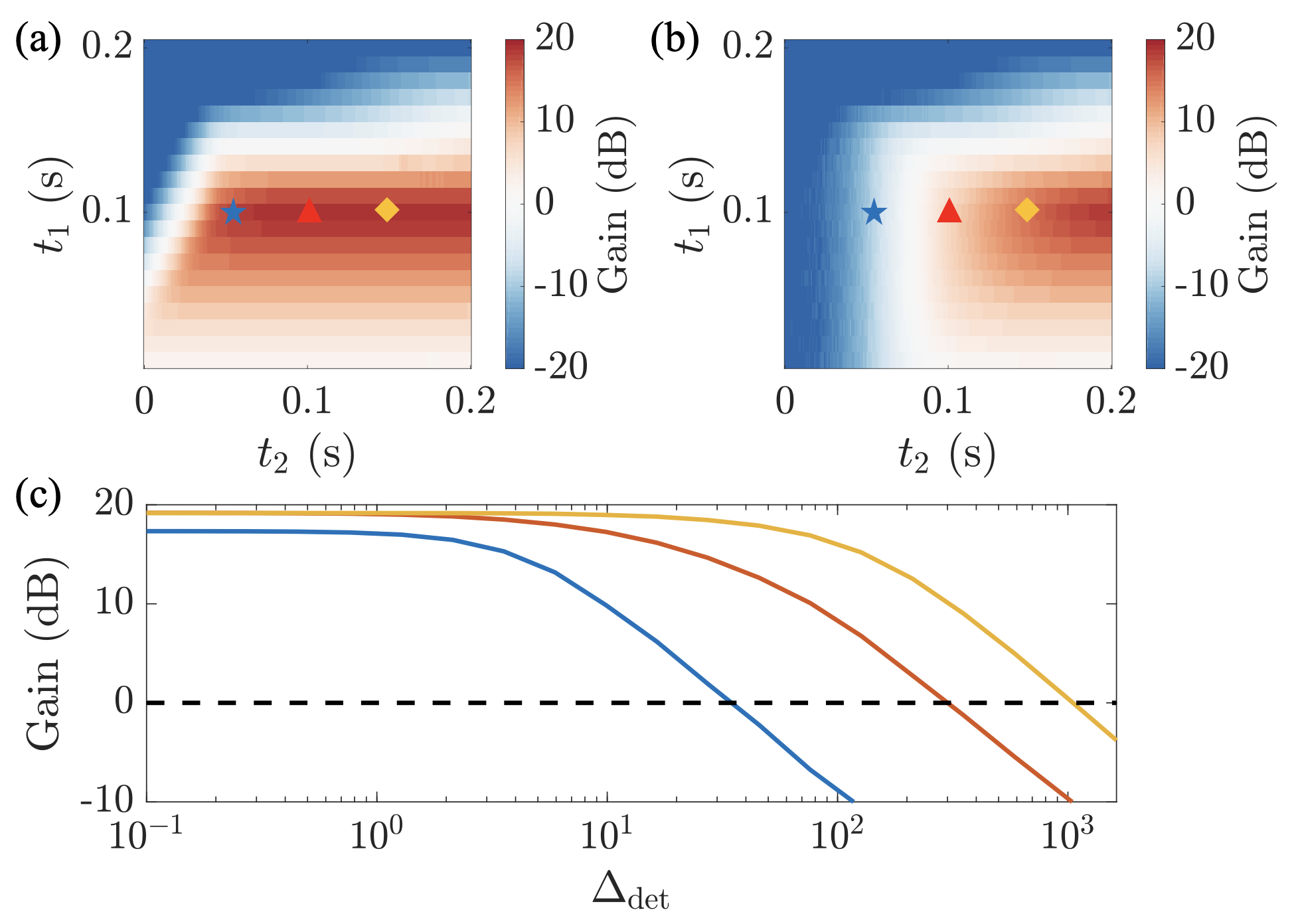}
	\caption{Robustness to detection noise. (a) Metrological gain without detection noise. (b) Metrological gain with detection noise of 100. (c) Effects of detection noise $\Delta_{\rm det}$ with $t_2=0.05~\rm{s}$ (blue line), $t_2=0.1~\rm{s}$ (red line) and $t_2=1.5~\rm{s}$ (yellow line).}
	\label{fig:detection}
\end{figure}

One major advantage of nonlinear readout protocol is robust to detection noise, which is quantitatively shown in Fig.~\ref{fig:detection}. For simplicity, we study the dependence of metrological gain on $t_1$, $t_2$, and detection noise $\Delta_{\rm{det}}$ by restricting the phase encoded to a small range of $\left[0,0.04\right]~\rm{rad}$. We find that in the absence of detection noise (Fig.~\ref{fig:detection}(a)), large gains can be obtained within a wide range of $t_2$. For short nonlinear splitting time $t_1<0.1~\rm{s}$ which gives rise to spin-nematic squeezed state, the gains nearly saturate right before the time-reversal point of $t_2=t_1$. However, the situation is different if we consider $\Delta_{\rm{det}}=100$ in Fig.~\ref{fig:detection}(b), which shows metrological gains above 0 are only maintained for $t_2\geq t_1$. In particular, such an over-reversed configuration gives rise to even higher gains than the perfect time-reversed configuration as demonstrated in Fig.~\ref{fig:detection}(c). This feature is in accordance with previous experiments in optical systems \cite{Manceau:2017uq} as well as in  spinor BECs \cite{Liu2021}. We attribute the enhanced robustness to the over amplified quantum noise $\Delta\rho_0$ as due to prolonged nonlinear recombining time, which causes the contribution of detection noise to be less dominant.

\subsection{Comparison between linear and nonlinear readout protocols}
Linear and nonlinear readout protocols provide different metrological gains, even with the same probe states employed for quantum sensing. In linear interferometry, entangled probe state suppresses quantum noise, while keeps the signal unchanged in reference  to the corresponding  classical probe state. Therefore the metrological gain is determined by the squeezing parameter. While for nonlinear readout, entangled probe state is disentangled before final detection to realize signal amplification. In this latter case, the gain remains mainly determined by the quantum Fisher information $F_Q$, or anti-squeezing for the squeezed probe state. The associated QCRB $(\Delta\phi)_{\rm{QCRB}}=1/\sqrt{F_Q}$ can be saturated at small encoded phase from the measurement of Loschmidt echo~\cite{Macri:2016aa}. In addition to improved robustness to detection noise, nonlinear readout protocol is also superior to the linear counterpart in fully exploiting  non-Gaussian entangled state for precision measurement, as demonstrated in two recent experiments~\cite{Liu2021, Colombo:2022tz}. For the spin-nematic nonlinear interferometry studied here, we work in the squeezed regime, the effective time-reversal can be achieved through probe state rotation instead of flipping the sign of Hamiltonian. Therefore, one would expect the two readout protocols discussed to achieve nearly the same metrological gains in the absence of noise, which is equivalent to the squeezing parameter.\par
\begin{figure}[t]
	\centering
	\includegraphics[width=1\linewidth]{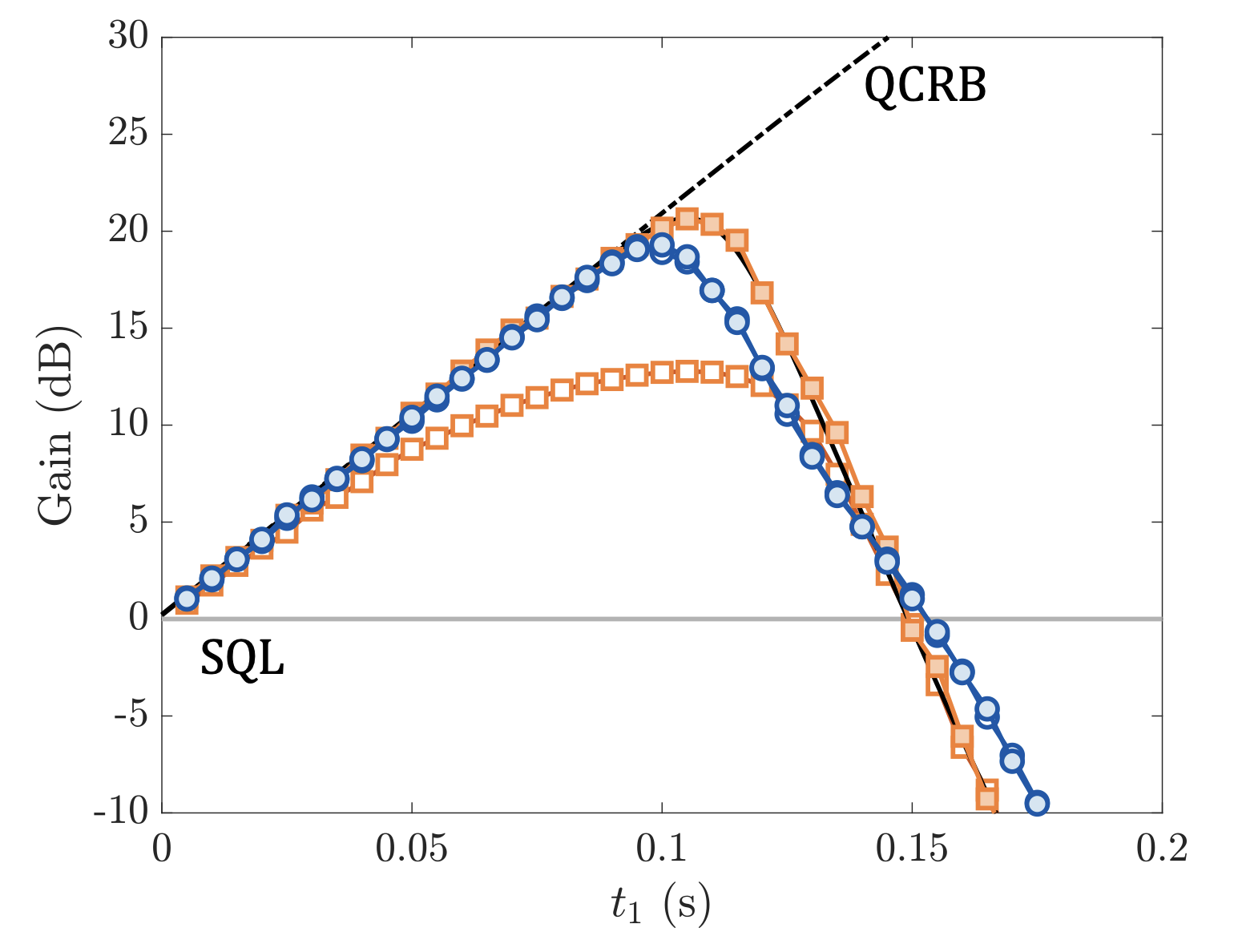}
	\caption{Metrological gains obtained with linear and nonlinear readout protocol. We optimize the gain over $t_2\in[0,0.2]~\rm{s}$ for each $t_1$. In the absence of noise, the sensitivity achieved by detecting the probe state with linear observable (orange solid squares) can saturate the squeezing factor (black solid line), while degrades with detection noise (orange open squares) of $\Delta_{\rm{det}}=24$. In contrast, the nonlinear readout protocol obtains slightly lower metrological gains than the squeezing parameters, but are insensitive to detection noise (solid and open blue dots). Black dot-dashed line stands for the QCRB.}
	\label{fig:lin_vs_non}
\end{figure}
The quantitative comparisons are shown in Fig.~\ref{fig:lin_vs_non}, where we calculate the gains for various nonlinear splitting time $t_1$ and optimize over nonlinear recombining time $t_2$. We find in the ideal case, the metrological gains achieved by linear readout (orange solid dots) always agree with the squeezing parameter (black solid line), while the nonlinear readout protocol can saturate the QCRB (black dashed line) before the instant of  optimal squeezing at $t_1=0.1~\rm{s}$, and declines in the over-squeezed regime. For all the probe states generated by nonlinear splitter, the optimal gains achieved in the nonlinear readout protocol are nearly the same or slightly worse than the values achieved with linear protocol. However, the two protocols show strikingly different responses to detection noise: the linear one (orange open dots) degrades with detection noise while the nonlinear one (blue open dots) essentially remains unaltered. 
%This feature explains the seemingly surprised results given in the main text, in which the gain achieved in nonlinear interferometer is higher than the directly measured squeezing parameter. 
%We also want to emphasize that the small discrepancy between linear and nonlinear readout protocol without detection noise is a compromise by the magnetic field noise, which forces us to select $\rho_0$ as the estimator. However, in principle, one can select the quadrature operator $\hat Q$ as a better estimator, which can saturate the optimal squeezing parameter. This will be the same for spin squeezed experiments which is implemented in clock states suffering less from magnetic field noise.

\subsection{Susceptibility to dynamical noises}
Besides detection noise, entanglement enhanced interferometers also suffer from various dynamical noises, such as those due to particle loss and decoherence due to fluctuating QZS. Typical two-mode spin squeezing experiments in BEC systems employ clock states composed of hyperfine ground and excited states. The spin dynamics then becomes immune to first-order magnetic field fluctuation, and condensate lifetime is limited by two-body or three-body inelastic collisions. Here we find the spin-nematic nonlinear interferometry is robust to particle loss and also robust fluctuating QZS.
%, and certain control errors of rotations in experiments.

\subsubsection{Atom loss}
\begin{figure}[t]
	\centering
	\includegraphics[width=1\linewidth]{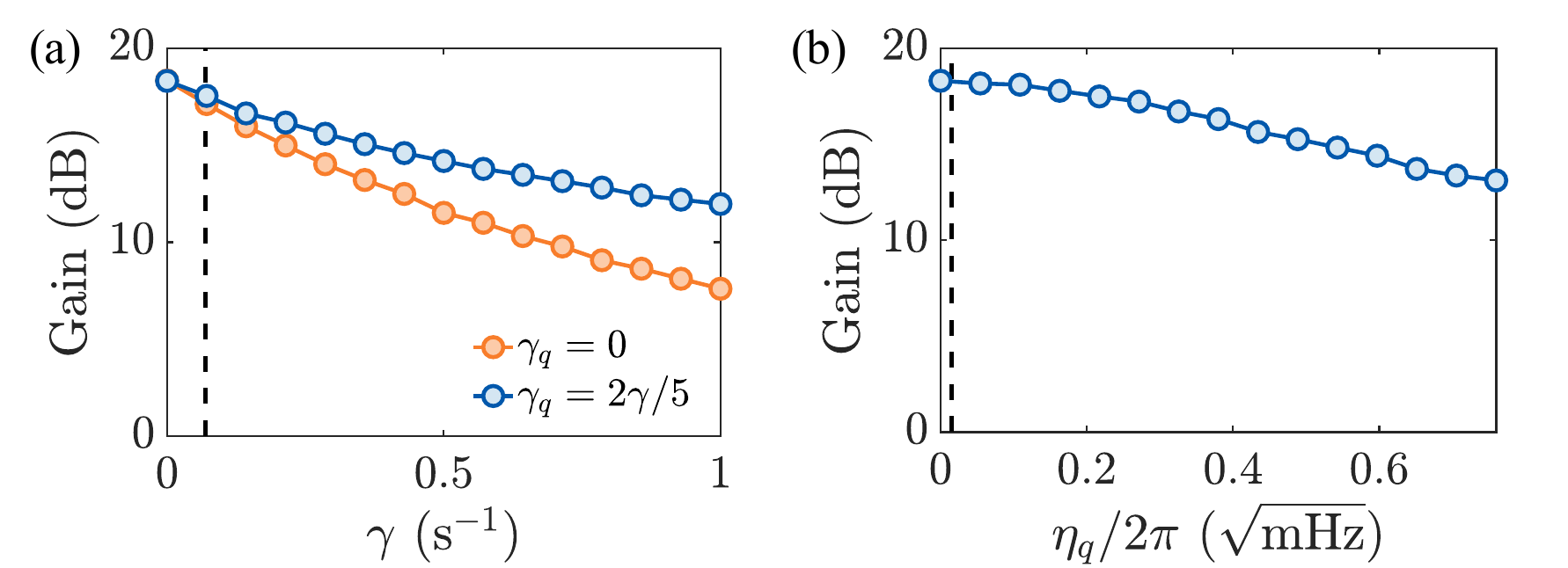}
	\caption{Influence of dynamical noises on metrological gain. The spin-nematic nonlinear interferometer is robust to both atom loss (a) and fluctuation of QZS (b). The deterioration by atom loss can be further mitigated by ramping QZS with a rate of $2\gamma/5$, such that the ratio $q(t)/c_2(t)$ remains fixed during evolution. Black dashed line denotes the experimentally calibrate noises level. The calculation is performed with $t_1=90~\rm{ms}$ and $t_2=129~\rm{ms}$.}
	\label{fig:noise}
\end{figure}

In our system, the lifetime of spinor BEC is mainly limited by three-body recombination in the $F=1$ hyperfine ground state. However, since the duration of the full spin-nematic interferometer is $0.21~\rm{s}$, nearly two orders of magnitude shorter than the condensate lifetime of $14.5~\rm{s}$, we expect a negligible influence of atom loss on the achieved metrological gain. This is quantitatively confirmed by Fig.~\ref{fig:noise}(a) which shows atom loss rate only degrades the gain by $1~\rm{dB}$ in our experiments. One can further compensate for this loss by ramping the QZS, such that the ratio of $c_2(t)/q(t)$ is fixed during the evolution, as adopted in our earlier work~\cite{Liu2021}. 

\subsubsection{Fluctuation of Zeeman shift}
Since the spin-mixing Hamiltonian Eq.~(\ref{SU11ham}) create $|\pm1\rangle$ atoms in pairs, the spin-nematic squeezed state is a superposition of Fock states with conserved magnetization. In this sense, the nonlinear interferometry is naturally implemented in a decoherence-free subspace, which is immune to the first-order of magnetic field fluctuation. Although the phase encoding process breaks the conservation of magnetization and leads occupations of states with nonzero magnetization, that do not interfere with each other during the subsequent nonlinear recombining. Therefore, if we choose $\hat{\rho}_0$ as the observable, which commutes with $\hat{L}_z$, the magnetization fluctuation plays no role in determining the metrological gain. In addition, thanks to the short-term evolution time in the undepleted pump regime, the fluctuation of QZS also barely deteriorates metrological gain as shown in Fig.~\ref{fig:noise}(b).
\section{Potential applications in atomic magnetometry}
The implemented spin-nematic nonlinear interferometers have several potential applications for quantum metrology. For example, the effectively time-reversed readout can be directly adopted in atomic clock to improve robustness to detection noise, which is one of the major limits in  an earlier study~\cite{Kruse}. Here, we focus on the application to atomic magnetometry, where atomic BEC is capable of mapping weak magnetic field with $\sim\mu m$ level spatial resolutions.

In Fig.~3 of the main text, we infer a sensing precision of MW magnetic field from the quadrature phase measurement. Here we provide more detailed explanations. During the phase encoding in the discussed nonlinear Ramsey interferometry, we turn on an off-resonant MW dressing field to encode a quadrature phase, which is proportional to QZS as defined by $q=(\varepsilon_1+\varepsilon_{-1})/2-\varepsilon_0$ with $\varepsilon_i$  the energy of the $|i\rangle~(i=0,\pm1)$ state. The resolution of $q$ is directly related to the phase resolution through $\Delta q=\Delta\phi/\tau$ where $\tau$ is the phase encoding time. For practical MW field sensing, we propose to apply a weak microwave field nearly resonant with the $|F=1,m_F=0\rangle$ to $|2,0\rangle$ clock transition. The effective MW Rabi frequency is given by
\begin{equation}
\Omega_{\rm{MW}}={g_J\mu_B\langle1,0|\hat{J}_z|2, 0\rangle}B=\mu_BB,
\end{equation}
with $g_J=2$ the fine structure Land\'e-g factor and $\hat{J}_z$ the $z$-component of electronic angular momentum. The coupling of MW magnetic field with atoms will induce an AC stark shift of $\Delta\varepsilon_0=\sqrt{\Delta^2+\Omega_{\rm{MW}}^2}/2$ to the energy of $|1,0\rangle$ state, which converges to $\Omega_{\rm{MW}}/2$ in the limit of $\Delta\rightarrow0$. Since the dressing to $|1,\pm1\rangle$ is negligible, we have a shifted QZS of $\Delta q=\Delta\varepsilon_0=\Omega_{\rm{MW}}/2$, which gives the single-shot magnetic field resolution of $B=2\Delta q/\mu_B$.

\clearpage
	
	\bibliography{ref}